\documentstyle[aps,prl,epsfig,amssymb]{revtex}

\begin{document}

\title{Deep-water internal solitary waves near critical density ratio }
\author{D.S. Agafontsev $^{a}$, F. Dias $^{b}$ and E.A. Kuznetsov $^{a}$}
\address{$^{a}$ L. D. Landau Institute for Theoretical Physics,
 2 Kosygin str., 119334 Moscow, Russia\\
 $^{b}$ Centre de Math\'ematiques et de Leurs Applications, 
 Ecole Normale Sup\'erieure de Cachan, \\ 61 avenue du
Pr\'esident Wilson,
 94235 Cachan cedex, France}
 
\maketitle
\begin{abstract}
Bifurcations of solitary waves propagating along the interface between two
ideal fluids are considered. The study is based on a Hamiltonian approach.
It concentrates on values of the density ratio close to a critical one,
where the supercritical bifurcation changes to the subcritical one. As the
solitary wave velocity approaches the minimum phase velocity of linear
interfacial waves (the bifurcation point), the solitary wave solutions
transform into envelope solitons. In order to describe their behavior and
bifurcations, a generalized nonlinear Schr\"{o}dinger equation describing
the behavior of solitons and their bifurcations is derived. In comparison
with the classical NLS equation this equation takes into account three
additional nonlinear terms: the so-called Lifshitz term responsible for
pulse steepening, a nonlocal term analogous to that first found by Dysthe
for gravity waves and the six-wave interaction term. We study both
analytically and numerically two solitary wave families of this equation for
values of the density ratio $\rho$ that are both above and below the
critical density ratio $\rho_{cr}$. For $\rho>\rho_{cr}$, the soliton
solution can be found explicitly at the bifurcation point. The maximum
amplitude of such a soliton is proportional to $\sqrt{\rho-\rho_{cr}}$, and
at large distances the soliton amplitude decays algebraically. A stability
analysis shows that solitons below the critical ratio are stable in the
Lyapunov sense in the wide range of soliton parameters. Above the critical
density ratio solitons are shown to be unstable with respect to finite
perturbations.\\

\end{abstract}





\section{Introduction}

The main goal of this paper is to study bifurcations for one-dimensional
internal solitary waves propagating along the interface between two ideal
fluids with different densities $\rho_1$ and $\rho_2$. The lighter fluid
with density $\rho_2$ lies above the heavier fluid with density $\rho_1$: $%
\rho=\rho_2/ \rho_1<1 $. These bifurcations occur if the solitary wave
velocity $V$ coincides with the minimum phase velocity $V_{cr}$ of linear
internal waves. If the upper density $\rho_2$ is small enough compared to
the lower density $\rho_1$, then a bifurcation similar to that for pure
gravity-capillary waves occurs \cite{LH89,IK,VD,DI93}. In this case solitary
waves undergo a supercritical bifurcation at the critical velocity: their
form approaches the form of the envelope solitons for the focusing
one-dimensional nonlinear Schr\"{o}dinger equation (1D NLSE) \cite
{LH93,Akylas}. The soliton amplitude behaves universally near the critical
velocity $V =V_{cr}$: it vanishes like $(V_{cr}-V)^{1/2}$. The width of the
solitary wave increases proportionally to $(V_{cr}-V)^{-1/2}$.

As the density ratio $\rho $ increases, the character of the nonlinear
interactions changes. The four-wave coupling coefficient decreases and
vanishes at $\rho =\rho _{cr}=(21-8\sqrt{5})/11$ \cite{DI96}, that is when $%
\rho \approx 0.283$. Such a value may not be obtained easily in a two-fluid
configuration. However, it may be relevant in a three-layer (or more)
configuration \cite{grue}, or in nonlinear optics, where similar
singularities can occur. Above the critical ratio, solitary waves undergo a
subcritical bifurcation: at the critical velocity, their amplitude jumps
from zero for $\rho $ below $\rho _{cr}$, up to finite values when $\rho $
is above the critical density ratio. In order to describe such type of
bifurcation, it is necessary to keep the next order terms beyond the
classical 1D NLSE. When the density ratio varies in a neighborhood of the
critical density ratio, it is possible to use, as in the derivation of the
classical NLSE, perturbation theory assuming that the interacting wave
amplitudes are small. At leading order, one needs to keep three kinds of
terms. The first one, coming from the four-wave interaction, takes into
account the so-called Lifshitz invariant \cite{LL}. This term is local
relative to the amplitude of the soliton and its first spatial derivative
and, as was shown in \cite{K99}, appears from the expansion of four-wave
interaction element under the assumption about its analytical dependence.
However, for deep-water internal waves, as we show in this paper, there
exists also a nonlocal term which has the same order of magnitude as the
first one \cite{sun}. Its structure is similar to the Dysthe term first
found for water (gravity) waves \cite{Dysthe} (see also \cite{Ablowitz2000}%
). In the case of water waves, this term is responsible for the interaction
of a narrow wave packet (in $k$-space) with mean flow, induced by the
packet. The third term takes into account six-wave interactions. In order to
find the six-wave coupling coefficient, one needs to calculate all possible
renormalizations due to three-, four- and five-wave interactions and
therefore this partial problem requires a lot of cumbersome calculations.
For these calculations we use the Hamiltonian formalism (see the review \cite
{ZK97}, as well as the papers \cite{ZK98,K99}), which appears to be the most
adequate method for this subject. In \cite{DI96}, a different method was
used: the problem was reformulated as a spatial dynamical system and only
the reversibility was exploited. It is necessary to underline also that the
use of the Hamiltonian approach to study solitary waves gives an appropriate
framework for the temporal behavior of the dynamics of solitary waves, e.g.
their stability. Through this formalism, it is easy to perform different
kinds of averaging and perturbations. Second, it is crucial that by applying
the Hamiltonian technique the averaging equations of motion retain their
original Hamiltonian form. In particular, this is of great help for the
investigation of soliton stability (see, for instance, \cite{K99}).

\section{Basic equations}

\setcounter{equation}{0}

Consider the interface $z=\eta (x,t)$ between two ideal incompressible
fluids with respective densities $\rho _1$ and $\rho _2$, in the presence of
gravity (with the acceleration $g$ acting down the vertical $z-$axis) and
capillarity with interfacial tension $\sigma $. We shall assume that the
lighter fluid with density $\rho _2$ occupies the region $\infty >z>\eta
(x,t)$, and respectively the heavier fluid occupies the region $-\infty
<z<\eta (x,t)$. Flows of both fluids are considered to be potential and
two-dimensional. The fluid velocities are given by
\[
\mathbf{v}_{1,2}=\nabla \phi _{1,2},
\]
where the velocity potentials $\phi _1$ and $\phi _2$ satisfy Laplace's
equation
\begin{equation}
\Delta \phi _{1,2}=0.  \label{laplace}
\end{equation}
These equations are subject to the following boundary conditions. Far from
the interface as $z\to \pm \infty $
\[
\phi _{1,2}\to 0.
\]
On the interface $z=\eta (x,t)$ the kinematic conditions hold:
\begin{equation}
\frac{\partial \eta }{\partial t}=(-v_x\eta _x+v_z)_{1,2}.  \label{kin}
\end{equation}
The dynamic condition reduces to the discontinuity of pressures across the
interface due to capillarity:
\[
p_1-p_2= -\sigma \frac \partial {\partial x}\left( \frac{\eta _x}{\sqrt{\eta
_x^2+1}}\right) .
\]
The use of Bernoulli equations in each fluid allows to rewrite the latter
equation in terms of potentials and their derivatives:
\begin{equation}
\rho_1\left( \frac{\partial \phi _2}{\partial t}+\frac 12(\nabla \phi
_2)^2+g\eta \right) -\rho_2\left( \frac{\partial \phi _1}{\partial t}+\frac
12(\nabla \phi _1)^2+g\eta \right) =\sigma \frac \partial {\partial x}\left(
\frac{\eta _x}{\sqrt{\eta _x^2+1}}\right) .  \label{bern}
\end{equation}
The equations (\ref{laplace})--(\ref{bern}) conserve the total energy:
\begin{equation}
H=K+U,  \label{ham}
\end{equation}
where the kinetic energy is equal to
\[
K=\int_{z>\eta }\frac{\rho _2(\nabla \phi _2)^2}2\;d\mathbf{r}+\int_{z<\eta }%
\frac{\rho _1(\nabla \phi _1)^2}2\;d\mathbf{r}
\]
and the potential energy is given by the expression
\[
U=\int (\rho _1-\rho _2)\frac{g\eta ^2}2\;dx+\int \sigma \left( \sqrt{\eta
_x^2+1}-1\right) \;dx.
\]
As shown first in \cite{Kontorovich} (see also \cite
{KS76,kuz-spec-zakh,ZK97,DB94}), the equations of motion (\ref{kin}) and (%
\ref{bern}) together with the Laplace equations (\ref{laplace}) represent a
Hamiltonian system. The Hamiltonian coincides with the energy (\ref{ham}).
The new variables $\Psi =(\rho _1\psi _1-\rho _2\psi _2)$ and the interface
shape $\eta $ are canonical conjugate variables:
\begin{equation}
\frac{\partial \eta }{\partial t}=\frac{\delta H}{\delta \Psi },\smallskip\
\ \frac{\partial \Psi }{\partial t}=-\frac{\delta H}{\delta \eta },
\label{canonical}
\end{equation}
where $\psi _{1,2}=\phi _{1,2}|_{z=\eta }$. The given Hamiltonian form
generalizes Zakharov's canonical form for free-surface hydrodynamics \cite
{Zakharov}. A Hamiltonian formulation of the problem of a free interface
between two ideal fluids, under rigid lid boundary conditions for the upper
fluid, was also given by Benjamin \& Bridges \cite{BB}. Craig \& Groves \cite
{CG} give a similar expression, by using the Dirichlet-Neumann operators for
both the upper and lower fluid domains (see also \cite{CGK}).

The Hamiltonian can be expanded in series with respect to powers of the
canonical variables. In this case the steepness of the interface plays the
role of a small parameter of expansion. Due to the conservation of the total
mass for each fluid this expansion begins with the quadratic term.

It is more convenient to work in Fourier space. Let us introduce the normal
variables $a_k$ by means of the following formulas:
\begin{eqnarray}  \label{normal-var}
\Psi(k)= i\sqrt{\frac{(1+\rho)\omega_k}{2|k|}}(a_k-a^*_{-k}), \\
\eta(k)= \sqrt{\frac{|k|}{2(1+\rho)\omega_k}}(a_k+a^*_{-k}),  \nonumber
\end{eqnarray}
where the density $\rho_1$ is set equal to unity and $\rho_2=\rho$. In these
formulas
\begin{equation}  \label{dispersion}
\omega_k=\left(\frac{|k|}{1+\rho}[g(1-\rho)+\sigma k^2]\right)^{1/2}
\end{equation}
is the dispersion relation for linear internal waves and $k$ is the wave
vector directed along the $x-$axis (1D case).

The transformation (\ref{normal-var}) diagonalizes the quadratic part of the
Hamiltonian,
\[
H_0=\int\omega_k |a_k|^2 dk.
\]
As a result, the equations of motion in the new variables $a_k$ take the
standard form \cite{ZK97}:
\begin{equation}  \label{k-rep}
\frac{\partial a_k}{\partial t}=-i\frac{\delta H}{\delta a^*_k},
\end{equation}
with the Hamiltonian
\[
H=H_0+H_{\mathrm{int}},
\]
where $H_{\mathrm{int}}$ is responsible for the nonlinear interactions
between waves. In the given case of internal waves the expansion of $H_{%
\mathrm{int}}$ in the wave amplitude will contain powers starting with the
cubic terms.

Consider a solution of Eq. (\ref{k-rep}) in the form of a solitary wave
propagating with constant velocity $V$. Then the potentials and the shape $%
\eta$ of the interface will depend on $x$ and $t$ through the combination $%
x-Vt$. In particular, the inverse Fourier transform of $a_k$,
\[
\psi(x,t)=\frac{1}{\sqrt{2\pi}} \int a_k(t)e^{ikx}dk,
\]
will be a function of $(x-Vt)$ only.

In Fourier space such dependence implies an exponential dependence in time
for the normal variables:
\[
a_k(t)=c_ke^{-ikVt},
\]
where the time-independent amplitude $c_k$ is defined from the equation
\begin{equation}  \label{sol}
(\omega _k-kV)\,c_k=-\frac{\delta H_{\mathrm{int}}}{\delta c_k^{*}}\equiv
f_k.
\end{equation}
This equation can be casted into the following variational problem:
\begin{equation}  \label{var}
\delta(H-VP)=0,
\end{equation}
where $P=\int k|c_k|^2 dk$ is the total momentum of the wave system. This
means that a solution to this equation represents a stationary point of the
Hamiltonian $H$ for fixed momentum $P$.

A solution to Eq. (\ref{sol}) in the form of a solitary wave is possible if
the difference $\omega _k- k V$ is sign-definite. When the equation
\begin{equation}  \label{cher}
\omega _k=k V
\end{equation}
has real roots (say $k=k_0$), then, in accordance with $x\delta(x)=0$, the
solution of (\ref{sol}) will be of the form:
\[
c_k=A_1\delta(k-k_0)+\frac{f_k}{\omega _k-kV}, \quad f_{k_0}=0.
\]
Hence taking the first term as the zero approximation, after iteration the
solution of this equation can be represented as an infinite series with
respect to $\delta(k-nk_0)$:
\[
c_k=\sum_n A_n\delta(k-nk_0).
\]
In $x-$space, this solution is equivalent to a periodic solution (for
details, see \cite{K99,ZK98}). Physically, this criterion is very
transparent. The equality (\ref{cher}) is the resonance condition for
Cherenkov radiation of waves by an object moving with the velocity $V$. Due
to such radiation a solitary wave will lose its energy and therefore cannot
be steady.

For the internal wave dispersion (\ref{dispersion}) the maximum solitary
wave velocity $V$ coincides with the minimum phase velocity of linear waves:
\[
V_{cr}=\min \,\frac{\omega _k}k.
\]
It occurs when
\begin{equation}
k=k_0=\left[ \frac{g(1-\rho )}\sigma \right] ^{1/2}.  \label{crit=k}
\end{equation}
At this point the values of the linear frequency and critical velocity are
\begin{equation}
\omega _0\equiv \omega (k_0)=\sqrt{2A gk_0}\quad \mbox{ and }\quad
V_{cr}\equiv \frac{\omega _0}{k_0}=\sqrt{\frac{2A g}{k_0}},
\label{crit-freq}
\end{equation}
where
\[
A=\frac{1-\rho }{1+\rho }
\]
is the Atwood number.

As the maximum solitary wave velocity is approached, the amplitude $c_k$
given by (\ref{sol}) reaches a very sharp maximum at the point $k=k_0$,
where the straight line $\omega =kV$ touches the dispersion curve $\omega
=\omega _k$:
\begin{equation}
c_k=\left[ \frac 12\omega ^{\prime \prime }\kappa ^2+k_0(V_{cr}-V)\right]
^{-1}f_k.  \label{res}
\end{equation}
Here $\kappa =k-k_0$ and $\omega ^{\prime \prime }=\partial ^2\omega
/\partial k^2$ is the positive-definite second derivative of $\omega _k$
taken at $k=k_0$: $\omega ^{\prime \prime }=\omega _0/(2k_0^2)>0$. Hence one
can see that as $V\to V_{cr}$ the width of the distribution $\Delta k$ tends
to zero, which corresponds to the peak at $k=k_0$ becoming narrower and
narrower. Due to the nonlinearities of the wave system this peak generates
multiple harmonics near $k=nk_0$ with integer $n$. If the amplitude of this
peak is small (if, for instance, it vanishes smoothly while approaching the
critical velocity) then we can use the perturbation theory that consists in
expanding $\psi $ through its harmonics:
\begin{equation}
\psi ({x^{\prime }})=\sum_{n=-\infty }^\infty \psi _n(X)e^{ink_0x^{\prime
}},\quad x^{\prime }=x-Vt.  \label{expand}
\end{equation}
Here the small parameter
\begin{equation}
\lambda =\sqrt{1-V/V_{cr}}  \label{lambda}
\end{equation}
and the ``slow'' coordinate $X=\lambda x^{\prime }$ are introduced, so that $%
\psi _n(X)$ is the amplitude of the envelope of \textit{n}-th harmonic. The
assumption that the solitary wave amplitude vanishes continuously at $%
V=V_{cr}$ means that the leading term of the series in Eq. (\ref{expand})
corresponds to the first harmonic, and all other harmonics are small with
respect to the parameter $\lambda $. This is the condition under which the
nonlinear Schr\"odinger equation is derived (see, for example, \cite
{ZK98,Z84,Newell}). In this case, at leading order in $\lambda $, we obtain
the stationary NLSE (compare with \cite{ZK98,K99,DK})
\begin{equation}
-\lambda ^2\omega _0\psi _1+\frac{\omega _0}{4k_0^2}\frac{\partial ^2\psi _1%
}{\partial x^2}-\mu |\psi _1|^2\psi _1=0,  \label{SNLS}
\end{equation}
where $\mu $ is related to the matrix element $\widetilde T_{k_0k_1k_2k_3}$
of four-wave interactions (see below) as
\begin{equation}
\mu =2\pi\widetilde T_{k_0k_0k_0k_0}.  \label{4wave0}
\end{equation}

>From now on, we drop the subscript $1$ for $\psi_1$. In complete
correspondence with (\ref{var}), the envelope equation (\ref{SNLS}) can be
recasted in the following variational problem:
\begin{equation}
\delta (\overline{H}+\omega_0\lambda ^2N)=0,  \label{var1}
\end{equation}
where
\[
N=\int |\psi |^2 \, dx
\]
is the number of waves or the wave action. The (averaged) Hamiltonian $%
\overline{H}$ is given by the expression
\[
\overline{H}=\frac 12\int \omega ^{\prime \prime }|\psi _x|^2 \; dx+H_{%
\mathrm{int}}^{(4)}.
\]
In this approximation the leading term in the interaction Hamiltonian $H_{%
\mathrm{int}}$ has the form
\begin{equation}
H_{\mathrm{int}}^{(4)}=\frac{\widetilde T_{k_0k_0k_0k_0}}2\int
c_k^{*}c_{k_1}^{*}c_{k_2}c_{k_3}\delta _{k+k_1-k_2-k_3}
\;dk_1dk_2dk_3dk_4=\frac \mu 2\int |\psi|^4\;dx.  \label{H40}
\end{equation}
The tilde denotes renormalization of the vertex $T$ due to the interaction
with the zeroth and second harmonics, corresponding to the cubic terms in
the Hamiltonian $H$. Thus, after averaging, the soliton solution is a
stationary point of the (averaged) Hamiltonian for fixed $N$.

>From Eq. (\ref{SNLS}) one can see that the localized solution is possible
if the coupling coefficient $\mu $ is negative (focusing nonlinearity). We
recall that in our case $\omega ^{\prime \prime }>0$. In this case Eq. (\ref
{SNLS}) can be rewritten in dimensionless variables as follows:
\begin{equation}
-\lambda ^2\psi +\psi _{xx}+|\psi |^2\psi =0.  \label{DNLS}
\end{equation}
Its soliton solution $\psi _s$ is given by
\begin{equation}
\psi _s=\frac{\sqrt{2}\lambda }{\cosh (\lambda x)}.  \label{NLSsoliton}
\end{equation}
Hence it follows that while approaching the critical velocity the soliton
amplitude vanishes like $\lambda =(1-V/V_{cr})^{1/2}$ and the soliton width
grows as $(1-V/V_{cr})^{-1/2}$. The latter means that our approximation
improves when approaching the critical velocity: the wave becomes more
monochromatic and nonlinearity weaker. This approximation becomes exact at
the critical velocity.

As we show in the next section such a situation occurs for all interfacial
solitary waves when the density ratio is less than the critical value $%
\rho_{cr}=(21-8\sqrt{5})/11$ (see for example \cite{DI96}). While increasing
$\rho$ the four-wave coupling coefficient $\mu$ remains negative up to the
critical ratio, where it vanishes. For $\rho>\rho_{cr}$ the coefficient $\mu$
becomes positive, so that the nonlinear interaction in (\ref{SNLS}) changes
its character, from focusing to defocusing. In this case, in order to have
solitary wave solutions, one needs to keep next order terms beyond the
classical nonlinear Schr\"odinger equation (\ref{SNLS}), which should
provide existence of localized solutions in the form of solitary waves. Such
solutions were computed numerically using the full water-wave equations by
Laget \& Dias \cite{LD}. Bridges et al. \cite{BCD} computed finite-amplitude
travelling waves near the transition from focusing to defocusing. The
simplest weakly nonlinear extension retains the terms due to six-wave
interactions. Such interactions should be of the focusing type in order to
compensate for the defocusing four-wave interaction. From this consideration
it becomes clear that the soliton amplitude undergoes a jump at the point $%
V=V_{cr}$. It is easy to estimate that such a jump will be proportional to $%
\sqrt{\mu}$. In order to obtain convergence of the Hamiltonian series
expansion, the jump must be small. In other words, such an expansion will be
valid if the deviation of $\rho$ from its critical value $\rho_{cr}$ is
small enough. The appearance of the jump at the critical velocity means that
the soliton undergoes subcritical bifurcation. Such type of bifurcation is
analogous to phase transition of first order. If the corresponding jump is
small then we have the analogue of the first-order phase transition close to
the second-order phase transition. For phase transitions such a situation
occurs in a small neighborhood of the so-called tri-critical point.

Now we will give the general structure of the Hamiltonian expansion
corresponding to interfacial waves near the critical density ratio assuming
the following two dimensionless parameters are small:
\[
\lambda =\sqrt{1-V/V_{cr}}\quad\mbox{and}\quad\theta=1-\rho /\rho _{cr}.
\]
As mentioned before, there are in this case three main contributions to
nonlinear terms (which, for instance, can balance dispersion, thus providing
the existence of stationary solitary waves). Two contributions come from the
expansion of the four-wave interaction Hamiltonian. Because a stationary
localized solution is assumed to be an envelope soliton, i.e. its spectrum
remains narrow and concentrated near $k=k_0$, we have to expand the
four-wave matrix element $\widetilde T_{k_1k_2k_3k_4}$, keeping the
first-order terms that are linear in $\kappa _i=k_i-k_0$. As shown in the
next section, this expansion contains local and nonlocal terms:
\begin{eqnarray}  \label{exprT}
\widetilde T_{k_1k_2k_3k_4} &=&\frac{\mu}{2\pi} +\frac{\beta}{2\pi} (\kappa
_1+\kappa _2+\kappa_3+\kappa _4) \\
&&\ -\frac{\gamma}{8\pi} (|\kappa _1-\kappa _3|+|\kappa _2-\kappa
_3|+|\kappa _2-\kappa _4|+|\kappa _1-\kappa _4|).  \nonumber
\end{eqnarray}
The constants $\beta$ and $\gamma$ have different parity relative to
reflection $k_0\to -k_0$. The coefficient $\beta $ changes its sign, but the
coefficient $\gamma $ retains its sign under this transform. The difference
in parities between $\beta$ and $\gamma$ gives different contributions to
the averaged four-wave Hamiltonian:
\begin{equation}
\overline{H}^{(4)}=\frac 12\int \left[ \mu |\psi |^4+2i\beta (\psi
_x^{*}\psi -\psi _x\psi ^{*})|\psi |^2 - \gamma |\psi |^2\widehat k |\psi|^2
\right] \;dx .  \label{wave4}
\end{equation}
Here $\widehat k$ is the positive definite integral operator
\[
\widehat k=-\partial _x\widehat H,
\]
and $\widehat H$ is the Hilbert transform:
\[
\widehat Hf(x)=\frac{1}{\pi} \left( P.V.\int_{-\infty }^\infty \frac{%
f(x^{\prime })dx^{\prime }}{x^{\prime }-x} \right).
\]
The Fourier transform of the kernel of the operator $\widehat k$ is equal to
$|k|$.

The third contribution, which is local in $\psi $, corresponds to six-wave
interactions:
\begin{equation}
\overline{H}^{(6)}=-C\int |\psi |^6 \;dx,  \label{6wave}
\end{equation}
where $C$ is the corresponding coupling coefficient.

The solitary wave shape in this case will be defined from the solution of
the following variational problem:
\begin{equation}
\delta (\overline{H}+\omega_0\lambda ^2 N)=0,  \label{var2}
\end{equation}
where the (averaged) Hamiltonian is given by the expression
\begin{equation}
\overline{H}=\frac 12\int \omega ^{\prime \prime }|\psi _x|^2dx+\overline{H}%
^{(4)}+\overline{H}^{(6)}.  \label{meanH}
\end{equation}
The terms $\overline{H}^{(4)}$ and $\overline{H}^{(6)}$ are defined by Eqs. (%
\ref{wave4}) and (\ref{6wave}), respectively. The variational problem (\ref
{var2}) can be considered as resulting from averaging the problem (\ref{var}%
) over `fast' spatial oscillations.

Thus, in order to solve the variational problem (\ref{var2}), we need to
know four coefficients: $\widetilde T_0 (=\widetilde T_{k_0k_0k_0k_0}),\beta
,\gamma $ and $C$. One can easily see that the contributions from terms
proportional to $\beta ,\gamma $ in $\overline{H}^{(4)}$ and the six-wave
Hamiltonian can be determined independently, which makes calculations more
simple.

\section{Hamiltonian expansion and matrix elements}

\setcounter{equation}{0}

We begin our calculations with the four-wave matrix element $\widetilde
T_{1234}$ in order to find $\widetilde{T}_0$ and its ``derivatives'' $\beta $
and $\gamma $.

The usual way to calculate matrix elements consists in expanding the
Hamiltonian in series with respect to powers of the canonical variables $%
\Psi $ and $\eta $. Then one substitutes in each term $H^{(n,m)}$ of the
Hamiltonian the variables $\Psi $ and $\eta $ expressed in terms of the
normal amplitudes $a_k$ and $a_k^{*}$ with the help of the formulas (\ref
{normal-var}), and finally one symmetrizes each term $H^{(n,m)}$ against all
$a_{k_i}$ and $a_{k_j}^{*}$. As a result one obtains
\[
H^{(n,m)} = \int
T_{k_1,...,k_n|k_{n+1},...,k_{n+m}}^{(n,m)}\prod_{i=1}^na_{k_i}%
\prod_{j=1}^ma_{k_{j+n}}^{*} \delta
(k_1+...+k_n-k_{n+1}-...-k_{n+m})\prod_{l=1}^{n+m}dk_l.
\]
The Hamiltonian for interactions $H_{\mathrm{int}}$ is represented as a sum
of $H^{(n,m)}$ terms:
\begin{equation}
H_{\mathrm{int}}=\sum_{n+m>2}H^{(n,m)},  \label{H-int}
\end{equation}
and the matrix elements $T_{k_1,...,k_n|k_{n+1},...,k_{n+m}}^{(n,m)}$ are
symmetric with respect to all permutations inside of both groups of indices $%
i=1,...,n$ and $j=n+1,...,n+m$. Moreover, $H^{(n,m)}=H^{*(m,n)}$.

To find the needed matrix elements, it is convenient to represent first the
Hamiltonian (\ref{ham}) in the following form by integration by parts:
\begin{equation}
H=\frac 12\int \left[ \mathcal{V}\Psi +(1-\rho )g\eta ^2+2\sigma \left(
\sqrt{1+\eta _x^2}-1\right) \right] \,dx,  \label{E:form24}
\end{equation}
where
\begin{equation}
\mathcal{V}\equiv V_{1,2}=\left( \frac{\partial \phi _{1,2}}{\partial z}%
-\eta _x\frac{\partial \phi _{1,2}}{\partial x}\right) _{z=\eta }.
\label{normal}
\end{equation}
Up to the multiplier $\sqrt{1+\eta _x^2}$, $\mathcal{V}$ coincides with the
normal component of the velocity $(\mathbf{v_{1,2}}\cdot \mathbf{n})$ on the
interface $z=\eta (x,t)$. The vector $\mathbf{n}=(1+\eta _x^2)^{-1/2}(-\eta
_x,1)$ is the unit normal to the interface.

Thus, only $\mathcal{V}$ needs to be expressed in terms of $\Psi $ and $\eta
$. To find this dependence we first solve the Laplace equations (\ref
{laplace}) for $\phi _1$ and $\phi _2$,
\begin{equation}
\phi _{1,2}(x,z)=\exp (\pm z\widehat{k})A_{1,2}(x),  \label{phi}
\end{equation}
where $A_{1,2}(x)$ are functions determined from the interface boundary
conditions and $\widehat{k}$ is the integral operator defined in the
previous section. The operator $E(z\widehat{k})=\exp (z\widehat{k})$ is
defined through the infinite series:
\begin{equation}
\exp (z\widehat{k})=1+z\widehat{k}+\frac 12z^2\widehat{k}^2+\frac 1{3!}z^3%
\widehat{k}^3+\cdots \;.  \label{exp}
\end{equation}
The boundary values $\psi _{1,2}$ of $\phi _{1,2}$ on the interface are
expressed by means of the operators $E(\pm \eta \widehat{k}):$
\[
\psi _{1,2}(x)=E(\pm \eta \widehat{k})A_{1,2}\equiv \exp [\pm \eta (x)%
\widehat{k}]A_{1,2}(x).
\]
Hence by calculating the derivatives of the potentials $\phi _{1,2}(x,z)$ at
the interface $z=\eta (x,t)$ we have the following expressions for $V_{1,2}$%
:
\[
\left( \frac{\partial \phi _{1,2}}{\partial z}-\eta _x\frac{\partial \phi
_{1,2}}{\partial x}\right) _{z=\eta }=\left\{ \pm E(\pm \eta \widehat{k}%
)\cdot \widehat{k}-\eta _xE(\pm \eta \widehat{k})\frac \partial {\partial
x}\right\} A_{1,2}(x).
\]
Writing down the equality between $V_1$ and $V_2$ yields
\begin{equation}  \label{V1=V2}
\left\{ \left( 1+\eta _x^2\right) E(\eta \widehat{k})\cdot \widehat{k}-\eta
_x\frac \partial {\partial x}E(\eta \widehat{k})\right\} A_1(x) =\left\{
-\left( 1+\eta _x^2\right) E(-\eta \widehat{k})\cdot \widehat{k}-\eta
_x\frac \partial {\partial x}E(-\eta \widehat{k})\right\} A_2(x),
\end{equation}
If in addition one uses the definition of $\Psi $,
\begin{equation}
\Psi =E(\eta \widehat{k})A_1(x)-\rho E(-\eta \widehat{k})A_2(x),
\label{psi-int}
\end{equation}
one has two relations to determine $\mathcal{V}$.

Let us introduce the operator
\begin{equation}
G=\left( 1+\eta _x^2\right) E(\eta \widehat{k})\cdot \widehat{k}\cdot E(\eta
\widehat{k})^{-1}-\eta _x\frac \partial {\partial x}.  \label{green}
\end{equation}
This operator represents the Green operator for one fluid (the lower fluid)
that establishes the relation between $V_1$ and $\psi _1$ on the surface $%
z=\eta (x)$:
\[
V_1=\left\{ \left( 1+\eta _x^2\right) E(\eta \widehat{k})\cdot \widehat{k}%
\cdot E(\eta \widehat{k})^{-1}-\eta _x\frac \partial {\partial x}\right\}
\psi _1(x).
\]
With the help of $G$, the kinetic energy of the lower fluid $K_1$ is
expressed as follows:
\[
K_1=\frac 12\int \psi _1G\psi _1 \, dx.
\]
This relation can be taken as the definition of the Green operator $G$ .
According to this definition this operator is self-adjoint as it should be.
Of course, this fact can be verified also by direct calculations, for
instance, by expanding $G$ with respect to powers of $\eta $. In this case
one needs first to expand the operator $E(\eta )^{-1}:$
\begin{eqnarray*}
E(\eta \widehat{k})^{-1} &=&1-\eta \widehat{k}+\eta \widehat{k}\eta \widehat{%
k}-\frac 12\eta ^2\widehat{k}^2 -\frac 1{3!}\eta ^3\widehat{k}^3+\frac
12\eta \widehat{k}\eta ^2\widehat{k}^2+\frac 12\eta ^2\widehat{k}^2\eta
\widehat{k}-\eta \widehat{k}\eta \widehat{k}\eta \widehat{k} \\
&&-\frac 1{4!}\eta ^4\widehat{k}^4+\frac 14\eta ^2\widehat{k}^2\eta ^2%
\widehat{k}^2+\frac 1{3!}\eta \widehat{k}\eta ^3\widehat{k}^3+\frac
1{3!}\eta ^3\widehat{k}^3\eta \widehat{k} \\
&&-\frac 12\eta ^2\widehat{k}^2\eta \widehat{k}\eta \widehat{k}-\frac 12\eta
\widehat{k}\eta ^2\widehat{k}^2\eta \widehat{k}-\frac 12\eta \widehat{k}\eta
\widehat{k}\eta ^2\widehat{k}^2+\eta \widehat{k}\eta \widehat{k}\eta
\widehat{k}\eta \widehat{k}+\cdots \,.
\end{eqnarray*}
Then the Green operator $G$ is written as a series in powers of $\eta $:
\begin{equation}
G=G^{(0)}+G^{(1)}+G^{(2)}+G^{(3)}+G^{(4)}+\cdots ,  \label{surf-green}
\end{equation}
where
\[
G^{(0)}=\widehat{k},\quad G^{(1)}=-\widehat{k}\eta \widehat{k}-\nabla \eta
\nabla ,\quad G^{(2)}=\widehat{k}\eta \widehat{k}\eta \widehat{k}-\frac 12%
\widehat{k}\left( \widehat{k}\eta ^2+\eta ^2\widehat{k}\right) \widehat{k},
\]
\[
G^{(3)} = \frac 12\widehat{k}\left( \widehat{k}\eta ^2\widehat{k}\eta +\eta
\widehat{k}\eta ^2\widehat{k}\right) \widehat{k}-\widehat{k}\eta \widehat{k}%
\eta \widehat{k}\eta \widehat{k} -\widehat{k}\left( \frac 13\nabla \eta
^3\nabla +\frac 12\eta ^2(\Delta \eta )\widehat{k}\right) -\frac 13\widehat{k%
}^2\eta ^3\widehat{k}^2,
\]
\begin{eqnarray*}
G^{(4)} &=&\widehat{k}\eta \widehat{k}\eta \widehat{k}\eta \widehat{k}\eta
\widehat{k}+\frac 14\widehat{k}^2\eta ^2\widehat{k}\eta ^2\widehat{k}^2+%
\widehat{k}\left( -\frac 12\eta ^2\widehat{k}^2\eta +\frac 16\eta ^3\widehat{%
k}^2\right) \widehat{k}\eta \widehat{k} \\
&&+\frac 14\widehat{k}^2\left( -\frac 16\widehat{k}^2\eta ^4+\eta ^2\widehat{%
k}^2\eta ^2\right) \widehat{k}+\frac 14\widehat{k}\left( -\frac 16\eta ^4%
\widehat{k}^2+\eta ^2\widehat{k}^2\eta ^2\right) \widehat{k}^2 \\
&&\ -\frac 12\widehat{k}\left( \widehat{k}\eta ^2\widehat{k}\eta \widehat{k}%
\eta +\eta \widehat{k}\eta \widehat{k}\eta ^2\widehat{k}\right) \widehat{k}+%
\widehat{k}\eta \widehat{k}\left( -\frac 12\eta ^2\widehat{k}^2\eta +\frac
16\eta ^3\widehat{k}^2\right) \widehat{k}.
\end{eqnarray*}

In the case of interfacial waves the Green operator $G_{\mathrm{in}}$,
defined by the relation $\mathcal{V}=G_{\mathrm{in}}\Psi $, is constructed
by solving the linear system (\ref{V1=V2},\ref{psi-int}) by means of the
operator $G$ (\ref{surf-green}) :
\begin{equation}
G_{\mathrm{in}}=\left( G^{-1}\left( \eta \right) +\rho G^{-1}(-\eta )\right)
^{-1}.  \label{green-in}
\end{equation}
Here the Green operator for the upper fluid $G_2$ is equal to $-G(-\eta )$.
The latter formula means that
\[
G_2=-G^{(0)}+G^{(1)}-G^{(2)}+G^{(3)}-G^{(4)}+\cdots \,.
\]
Hence one can see that the Green operator $G_{\mathrm{in}}$ is self-adjoint,
as it should be. The total kinetic energy of two fluids is defined by the
operator $G_{\mathrm{in}}:$%
\[
K=\frac 12\int \Psi G_{\mathrm{in}}\Psi dx.
\]
The expansion of the Green operator $G_{\mathrm{in}}$ in powers of $\eta $
is expressed through the expansion of $G\left( \eta \right) ^{-1}$:
\[
G\left( \eta \right) ^{-1}=\Gamma ^{(0)}+\Gamma ^{(1)}+\Gamma ^{(2)}+\Gamma
^{(3)}+\Gamma ^{(4)}+\cdots ,
\]
where
\[
\Gamma ^{(0)}=\widehat{k}^{-1},\quad \Gamma ^{(1)}=-\widehat{k}^{-1}G^{(1)}%
\widehat{k}^{-1},\quad \Gamma ^{(2)}=\widehat{k}^{-1}\left( G^{(1)}\widehat{k%
}^{-1}G^{(1)}-G^{(2)}\right) \widehat{k}^{-1}
\]
\[
\Gamma ^{(3)}=\widehat{k}^{-1}\left( -G^{(1)}\widehat{k}^{-1}G^{(1)}\widehat{%
k}^{-1}G^{(1)}-G^{(3)}+G^{(1)}\widehat{k}^{-1}G^{(2)}+G^{(2)}\widehat{k}%
^{-1}G^{(1)}\right) \widehat{k}^{-1}
\]
\begin{eqnarray*}
\Gamma ^{(4)} &=&\widehat{k}^{-1}(-G^{(4)}+G^{(1)}\widehat{k}%
^{-1}G^{(3)}+G^{(3)}\widehat{k}^{-1}G^{(1)} \\
&&+G^{(2)}\widehat{k}^{-1}G^{(2)}-G^{(2)}\widehat{k}^{-1}G^{(1)}\widehat{k}%
^{-1}G^{(1)}-G^{(1)}\widehat{k}^{-1}G^{(2)}\widehat{k}^{-1}G^{(1)} \\
&&\ -G^{(1)}\widehat{k}^{-1}G^{(1)}\widehat{k}^{-1}G^{(2)}+G^{(1)}\widehat{k}%
^{-1}G^{(1)}\widehat{k}^{-1}G^{(1)}\widehat{k}^{-1}G^{(1)})\widehat{k}^{-1}.
\end{eqnarray*}
The inverse Green operator $G_{\mathrm{in}}^{-1}$ is the combination
\[
G^{-1}\left( \eta \right) +\rho G^{-1}(-\eta )=\left( \rho +1\right) (\Gamma
^{(0)}+A\Gamma ^{(1)}+\Gamma ^{(2)}+A\Gamma ^{(3)}+\Gamma ^{(4)}+\cdots).
\]
Hence the Green operator $G_{\mathrm{in}}$ is given through the following
expansion:
\[
G_{\mathrm{in}}=G_{\mathrm{in}}^{(0)}+G_{\mathrm{in}}^{(1)}+G_{\mathrm{in}%
}^{(2)}+G_{\mathrm{in}}^{(3)}+G_{\mathrm{in}}^{(4)}+\cdots ,
\]
where
\begin{eqnarray}
G_{\mathrm{in}}^{(0)} &=&\left( \rho +1\right) ^{-1}\widehat{k},\quad G_{%
\mathrm{in}}^{(1)}=\left( \rho +1\right) ^{-1}AG^{(1)},  \label{G0-2} \\
\ G_{\mathrm{in}}^{(2)} &=&\left( \rho +1\right) ^{-1}\left(
G^{(2)}+(A^2-1)G^{(1)}\widehat{k}^{-1}G^{(1)}\right) ,
\end{eqnarray}
\begin{equation}
G_{\mathrm{in}}^{(3)}=A\left( \rho +1\right) ^{-1}\left[
G^{(3)}+(A^2-1)G^{(1)}\widehat{k}^{-1}G^{(1)}\widehat{k}^{-1}G^{(1)}\right] ,
\label{G3}
\end{equation}
\begin{eqnarray}  \label{G4}
G_{\mathrm{in}}^{(4)} &=&\left( \rho +1\right) ^{-1}\left[G^{(4)}-A^2G^{(1)}%
\widehat{k}^{-1}G^{(1)}\widehat{k}^{-1}G^{(1)}\widehat{k}^{-1}G^{(1)} \right.
\\
&& \left. +\left( A^2-1\right) \left( G^{(3)}\widehat{k}^{-1}G^{(1)}+G^{(1)}%
\widehat{k}^{-1}G^{(3)}-G^{(1)}\widehat{k}^{-1}G^{(2)}\widehat{k}%
^{-1}G^{(1)}\right) \right].  \nonumber
\end{eqnarray}
This expansion of $G_{\mathrm{in}}$ allows one to write down the expansion
for the Hamiltonian:

\[
H=H_0+H^{(3)}+H^{(4)}+H^{(5)}+H^{(6)}+\cdots,
\]
where
\begin{equation}
H_0=\int \left[ \frac{\Psi \widehat{k}\Psi }{2\left( \rho +1\right) }%
+(1-\rho )\frac{g\eta ^2}2+\sigma \frac{\eta _x^2}2\right] dx,  \label{H-0}
\end{equation}
\begin{equation}
H^{(3)}=\int \frac{A\eta }{2\left( \rho +1\right) }\left[ \Psi _x^2-\left(
\widehat{k}\Psi \right) ^2\right] dx,  \label{H-3}
\end{equation}
\begin{eqnarray}  \label{H-4}
H^{(4)} &=&\ \frac 1{2\left( \rho +1\right) }\int \Psi \left[A^2\widehat{k}%
\eta \widehat{k}\eta \widehat{k}-\frac 12\widehat{k}\left( \widehat{k}\eta
^2+\eta ^2\widehat{k}\right) \widehat{k} \right. \\
&&\ \ \left. +(A^2-1)\left( \partial _x\eta \partial _x\eta \widehat{k}+%
\widehat{k}\eta \partial _x\eta \partial _x-\partial _x\eta \widehat{k}\eta
\partial _x\right) \right]\Psi dx-\int \frac{\sigma \eta _x^4}8dx.  \nonumber
\end{eqnarray}
The Hamiltonians $H^{(5)}$ and $H^{(6)}$ are expressed respectively through $%
G_{\mathrm{in}}^{(3)}$ and $G_{\mathrm{in}}^{(4)}$ (\ref{G3},\ref{G4}):
\begin{equation}
H^{(5)}=\frac 12\int \Psi G_{\mathrm{in}}^{(3)}\Psi dx,  \label{H-5}
\end{equation}
\begin{equation}
H^{(6)}=\frac 12\int \Psi G_{\mathrm{in}}^{(4)}\Psi dx+\int \frac{\sigma
\eta _x^6}{16}dx.  \label{H-6}
\end{equation}

Note that the Hamiltonian $H^{(3)}$ for the interfacial case was first
calculated in the paper \cite{kuz-spec-zakh}. For surface waves ($\rho =0$
or $A=1$) the expression for $H^{(4)}$ in the form (\ref{H-4}) was presented
in the papers \cite{Zakharov}, \cite{zakh-dyach-kor}.

After these calculations we can find the needed matrix elements. First, let
us consider the Hamiltonian expansion in Fourier space (\ref{H-0}--\ref{H-6}%
). For $H_0$ and $H^{(3)}$ it gives
\begin{equation}
H_0=\frac 12\int \left\{ \frac k{1+\rho }\mid \psi _k\mid ^2+\left[ (1-\rho
)g+\sigma k^2\right] \mid \eta _k\mid ^2\right\} dk,  \label{HF-0}
\end{equation}
\begin{equation}
H^{(3)}=-\int \frac A{2\left( \rho +1\right) }\left[
|k_1||k_2|+k_1k_2\right] \Psi _1\Psi _2\eta _3\delta _{1+2+3}\frac{dk_{123}}{%
\sqrt{2\pi }}.  \label{HF-3}
\end{equation}
Here the subscript $1$ in $\Psi _1$ means $k_1$ and so on, while $%
dk_{123}\equiv dk_1dk_2dk_3$. With this notation the four-wave Hamiltonian
takes the form

\begin{equation}  \label{HF-4}
H^{(4)} = \int C_{12|34}\Psi _1\Psi _2\eta _3\eta _4\delta _{1+2+3+4}\frac{%
dk_{1234}}{(\sqrt{2\pi })^2} -\frac \sigma 8\int k_1k_2k_3k_4\eta _1\eta
_2\eta _3\eta _4\delta _{1+2+3+4}\frac{dk_{1234}}{(\sqrt{2\pi })^2},
\end{equation}
where
\begin{eqnarray*}
C_{12|34} &=&\frac 1{4\left( \rho +1\right) }\left\{\frac{A^2}2\left|
k_1\right| \left| k_2\right| \left( \left| k_1+k_3\right| +\left|
k_1+k_4\right| +\left| k_2+k_3\right| +\left| k_2+k_4\right| \right) \right.
\\
&&-\frac{A^2}2\left| k_1\right| \left| k_2\right| \left( \left| k_1\right|
+\left| k_2\right| \right) +(A^2-1)(\left| k_2\right| +\left| k_1\right|
)(k_1k_2-\left| k_1\right| |k_2|)- \\
&& \left. -\frac 12(A^2-1)k_1k_2\left( \left| k_1+k_3\right| +\left|
k_1+k_4\right| +\left| k_2+k_3\right| +\left| k_2+k_4\right| \right)
\right\}.
\end{eqnarray*}

Hence it can be checked easily that the transition to the normal variables
by means of (\ref{normal}) diagonalizes the quadratic Hamiltonian:
\[
H_0=\int \omega _k|a_k|^2dk,
\]
so that the equations of motion are written in the standard form (\ref{k-rep}%
).

Substituting now the transformation (\ref{normal}) into (\ref{HF-3}) gives
the following expression for $H^{(3)}$ :

\[
H^{(3)}=\frac 13\int U_{123}(a_1^{*}a_2^{*}a_3^{*}+a_1a_2a_3)\delta
_{1+2+3}dk_{123} +\int V_{1|23}(a_1^{*}a_2a_3+a_1a_2^{*}a_3^{*})\delta
_{1-2-3}dk_{123}.
\]
Here the matrix elements $U_{123}$ and $V_{1|23}$ are:
\begin{eqnarray}
\ \sqrt{2\pi }U_{123} &=&\frac A{4(1+\rho )^{1/2}}\left\{\left( \frac{%
k_3\omega _1\omega _2}{2k_1k_2\omega _3}\right) ^{1/2}\left[ k_1k_2+
|k_1||k_2| \right] \right.  \label{U} \\
&&\ \left. +\left( \frac{k_1\omega _2\omega _3}{2k_2k_3\omega _1}\right)
^{1/2}\left[ |k_2||k_3|+k_2k_3\right] +\left( \frac{k_2\omega _3\omega _1}{%
2k_3k_1\omega _2}\right) ^{1/2}\left[ |k_3||k_1|+k_3k_1\right] \right\},
\nonumber
\end{eqnarray}
\begin{eqnarray}
\ \sqrt{2\pi }V_{1|23} &=&\frac A{4(1+\rho )^{1/2}}\left\{\left( \frac{%
k_3\omega _1\omega _2}{2k_1k_2\omega _3}\right) ^{1/2}\left[
-|k_1||k_2|+k_1k_2\right] \right.  \label{V} \\
&&\ \left. +\left( \frac{k_1\omega _2\omega _3}{2k_2k_3\omega _1}\right)
^{1/2}\left[ |k_2||k_3|+k_2k_3\right] +\left( \frac{k_2\omega _3\omega _1}{%
2k_3k_1\omega _2}\right) ^{1/2}\left[ -|k_3||k_1|+k_3k_1\right] \right\}.
\nonumber
\end{eqnarray}
The Hamiltonian describing $2\rightarrow 2$ interacting waves has the form
\begin{eqnarray*}
H^{(2,2)}=\frac 12\int T_{12|34}a_1^{*}a_2^{*}a_3a_4\delta
_{1+2-3-4}dk_{1234}.
\end{eqnarray*}
Here the primitive (non-renormalized) 4-wave matrix element $%
T_{12|34}^{(2,2)}$ is given by the expression
\begin{eqnarray}
\left( \sqrt{2\pi }\right) ^2T_{12|34}
&=&2(R_{-1-2|34}+R_{34|-1-2}-R_{-13|-24}-R_{-14|-23}  \label{T-4} \\
&&\ \ \ \ \ \ \ \ \
-R_{-24|-13}-R_{-23|-14}+2P_{12|34}+2P_{13|24}+2P_{23|14}),  \nonumber
\end{eqnarray}
where
\begin{eqnarray*}
R_{12|34} &=&-\frac 14\left( \frac{\omega _1\omega _2k_3k_4}{\omega _3\omega
_4k_1k_2}\right) ^{1/2}C_{12|34}, \\
P_{12|34} &=&-\frac \sigma {32(1+\rho )^2}\left( \frac{k_1k_2k_3k_4}{\omega
_1\omega _2\omega _3\omega _4}\right) ^{1/2}k_1k_2k_3k_4.
\end{eqnarray*}
In this section we presented only three matrix elements: $U_{123}$, $V_{1|23}
$ and $T_{12|34}$. These matrix elements can be used not only for
one-dimensional surface waves, but also after obvious generalizations to the
general (2D) case also. All the other matrix elements can be found by the
same procedure as the one used to find $U_{123}$, $V_{1|23}$ and $T_{12|34}$.

\section{Calculations of coupling coefficients}

\setcounter{equation}{0}

However, for solitons near the bifurcation velocity $V=V_{cr}$ when the
density ratio is close to the critical one, the complete knowledge of $%
U_{123}$, $V_{1|23}$ and $T_{12|34}$ is sufficient. Indeed we only need to
know the coupling coefficients $\mu $, $\beta ,$ $\gamma $ and $C$ in (\ref
{wave4}) and (\ref{6wave}). A straightforward and classical way to find them
is to use the diagram technique \cite{Zakh-Lvov} based on the
renormalization of matrix elements. We will use this approach partially,
only to calculate the constant $\mu$. The three other coefficients ($\beta $%
, $\gamma $ and $C$) can be found by using the procedure of averaging with
respect to high frequencies. In the present case, it is the carrying
frequency $\omega _0$ of the main harmonic. The amplitudes of all the other
harmonics are assumed to be small in order to apply the perturbation
technique based on the Hamiltonian expansion.

First we compute the coefficient $\mu $. As is well known (see \cite{ZK98}
for example), if three-wave interactions are not resonant, they can be
excluded by canonical transformations that result in the renormalization of
the high-order matrix elements. In particular, for $2\rightarrow 2$
interacting waves, such a renormalization yields
\begin{eqnarray}
\widetilde T_{12|34} & = & T_{12|34}  \label{matrix9} \\
&&-V_{4|2,4-2}V_{1|3,1-3}\left( \frac 1{\omega _{4-2}+\omega _2-\omega
_4}+\frac 1{\omega _{1-3}+\omega _3-\omega _1}\right)  \nonumber \\
&&-V_{2|4,2-4}V_{3|1,3-1}\left( \frac 1{\omega _{2-4}+\omega _4-\omega
_2}+\frac 1{\omega _{3-1}+\omega _1-\omega _3}\right)  \nonumber \\
&&-V_{3|2,3-2}V_{1|4,1-4}\left( \frac 1{\omega _{3-2}+\omega _2-\omega
_3}+\frac 1{\omega _{1-4}+\omega _4-\omega _1}\right)  \nonumber \\
&&-V_{2|3,2-3}V_{4|1,4-1}\left( \frac 1{\omega _{2-3}+\omega _3-\omega
_2}+\frac 1{\omega _{4-1}+\omega _1-\omega _4}\right)  \nonumber \\
&&-V_{3+4|3,4}V_{1+2|1,2}\left( \frac 1{\omega _{3+4}-\omega _3-\omega
_4}+\frac 1{\omega _{1+2}-\omega _1-\omega _2}\right)  \nonumber \\
&&-U_{-3-4,3,4}U_{-1-2,1,2}\left( \frac 1{\omega _{3+4}+\omega _3+\omega
_4}+\frac 1{\omega _{1+2}+\omega _1+\omega _2}\right) .  \nonumber
\end{eqnarray}
Thus, in the four-wave interaction vertex, there are three contributions:
the first one comes from $G_{\mathrm{in}}^{(2)}$ ($\sim R$), the second
contribution is connected with capillarity ($\sim P$), and the last
contribution stems from the renormalization (\ref{matrix9}) due to
three-wave interactions. The latter written for $k_1=k_2=k_3=k_4=k_0$ ($%
\widetilde T_{k_0k_0|k_0k_0}\equiv \widetilde T_0$) represents the
interactions with the zeroth and second harmonics. However the interaction
with the zeroth harmonic vanishes because the three-wave matrix element $%
V_{1|23}$ tends to zero sufficiently rapidly if one of the wavenumbers $k_i$
tends to zero. Therefore
\begin{equation}
\widetilde T_0=T_{k_0k_0|k_0k_0}+\frac{2V_{2k_0|k_0k_0}^2}{2\omega
_{k_0}-\omega _{2k_0}}-\frac{2U_{k_0k_0-2k_0}^2}{2\omega _{k_0}+\omega
_{2k_0}}.  \label{crit-ratio}
\end{equation}
From (\ref{U}), (\ref{V}) and (\ref{T-4}) one obtains
\[
\ \sqrt{2\pi }U_{k_0k_0-2k_0}=\frac{Ak_0\omega _{k_0}}{2(1+\rho )^{1/2}}%
\left( \frac{k_0}{\omega _{2k_0}}\right) ^{1/2},
\]
\[
\ \sqrt{2\pi }V_{2k_0|k_0k_0}=\frac{Ak_0\omega _{k_0}}{2(1+\rho )^{1/2}}%
\left( \frac{k_0}{\omega _{2k_0}}\right) ^{1/2},
\]
\[
\left( \sqrt{2\pi }\right) ^2T_{k_0k_0|k_0k_0}=\frac 5{16}\frac{k_0^3}{%
\left( \rho +1\right) }.
\]
Substituting these expressions into (\ref{crit-ratio}) gives
\[
\widetilde T_0=\frac{k_0^3}{2\pi (1+\rho )}\left( A_{cr}^2-A^2\right) \equiv
\frac \mu {2\pi },
\]
where the square of the critical Atwood number $A_{cr}^2$ is equal to $5/16$%
. This gives for the critical value of $\rho $
\[
\rho _{cr}=\frac{4-\sqrt{5}}{4+\sqrt{5}},
\]
in agreement with the paper \cite{DI96}. For $\rho <\rho _{cr}$, the
four-wave coupling coefficient is negative, and the corresponding
nonlinearity is of the focusing type. In this case, solitary waves near the
critical velocity $V_{cr}$ are described by the stationary NLSE (\ref{SNLS})
and undergo a supercritical bifurcation at $V=V_{cr}$ \cite{DI96}. For $\rho
>\rho _{cr}$ the coupling coefficient changes sign and the bifurcation
becomes subcritical. To find the soliton shape in this case we need to
calculate three more coefficients: $\beta ,$ $\gamma $ and $C$. The first
two coefficients are defined from the expansion of $\widetilde T_{12|34}$
near $k_i=k_0$.

To find the coefficient $\gamma $ we shall use the averaging procedure.
There are two contributions to $\gamma .$ The first one comes from the
Hamiltonian (\ref{H-4}). Hence one can see that nonlocality arises from two
terms in $\overline{H}^{(4)}$:
\begin{equation}
\overline{H}_{\mathrm{nonlocal}}^{(4)}=\int \frac{k_0^2}{2\left( \rho
+1\right) }A^2\left\langle \Psi \eta \right\rangle \widehat{k}\left\langle
\Psi \eta \right\rangle \ dx+\int \frac 1{2\left( \rho +1\right)
}(A^2-1)\left\langle \partial \Psi \cdot \eta \right\rangle \widehat{k}%
\left\langle \partial \Psi \cdot \eta \right\rangle dx.  \label{H-NL}
\end{equation}
Here the brackets $\left\langle \cdots \right\rangle $ denote average with
respect to short-wave oscillations so that $\left\langle \Psi \eta
\right\rangle =0$ but $\left\langle \partial \Psi \cdot \eta \right\rangle
\neq 0$. The latter follows after substituting $\eta $ and $\Psi $ expressed
in terms of the envelope amplitude $\psi $:
\begin{eqnarray}
\eta &=&\left[ \frac{k_0}{2(1+\rho )\omega _0}\right] ^{1/2}\left( \psi
e^{-i\omega _0t+ik_0x}+\psi ^{*}e^{i\omega _0t-ik_0x}\right) ,
\label{normal1} \\
\Psi &=&-i\left[ \frac{(1+\rho )\omega _0}{2k_0}\right] ^{1/2}\left( \psi
e^{-i\omega _0t+ik_0x}-\psi ^{*}e^{i\omega _0t-ik_0x}\right).  \nonumber
\end{eqnarray}
It is interesting to note that these approximations of the exact formulas (%
\ref{normal}) have an accuracy of $(\Delta k/k_0)^2,$ where $\Delta k$ is
the width of the main peak, since at the critical wavenumber $k=k_0$
\[
\frac \partial {\partial k}\left( \frac{\omega _k}k\right) =0.
\]
In particular, computing the averages $\left\langle \Psi \eta \right\rangle $
and $\left\langle \partial \Psi \cdot \eta \right\rangle $ from (\ref
{normal1}) gives zero for the first mean and
\[
\left\langle \partial \Psi \cdot \eta \right\rangle =k_0|\psi |^2.
\]
As a result, the nonlocal Hamiltonian (\ref{H-NL}) has the form
\begin{equation}
\overline{H}_{\mathrm{nonlocal}}^{(4)}=\int \frac{k_0^2}{2\left( \rho
+1\right) }(A^2-1)|\psi |^2\widehat{k}|\psi |^2dx.  \label{NL-4}
\end{equation}
It is important that $\overline{H}_{\mathrm{nonlocal}}^{(4)}$ is negative
definite. This means that this term provides focusing.

Another contribution to $\gamma $ comes from the interaction with the zeroth
harmonic, which for gravity waves is responsible for the mean flow induced
by the wave packet. The same statement is also valid for interfacial waves.

Consider the three-wave Hamiltonian (\ref{H-3}) and average with respect to
fast oscillations with frequency $\omega _0$, under the assumption that $%
\eta $ and $\Psi $ contain two parts:
\begin{eqnarray}
\eta  &=&\left[ \frac{k_0}{2(1+\rho )\omega _0}\right] ^{1/2}\left( \psi
e^{-i\omega _0t+ik_0x}+\psi ^{*}e^{i\omega _0t-ik_0x}\right) +\widetilde{%
\eta },  \label{separation} \\
\Psi  &=&-i\left[ \frac{(1+\rho )\omega _0}{2k_0}\right] ^{1/2}\left( \psi
e^{-i\omega _0t+ik_0x}-\psi ^{*}e^{i\omega _0t-ik_0x}\right) +\widetilde{%
\Psi }.  \nonumber
\end{eqnarray}
In (\ref{separation}), $\widetilde{\eta }$ and $\widetilde{\Psi }$ are
low-frequency quantities responsible for the mean flow induced by the
high-frequency wave packet, concentrated at $k=k_0$. Substituting (\ref
{separation}) into the three-wave Hamiltonian (\ref{H-3}) and then averaging
yields
\begin{equation}
\overline{H}^{(3)}=\int \frac{A\left\langle \partial \Psi \cdot \eta
\right\rangle }{\left( \rho +1\right) }\widetilde{\Psi }_xdx=\int \frac{Ak_0%
}{\left( \rho +1\right) }|\psi |^2\widetilde{\Psi }_xdx.  \label{meanH-3}
\end{equation}
This Hamiltonian describes the interaction of the wave packet with the mean
flow $\partial \widetilde{\Psi }/\partial x$. To complete the system one
needs to add the quadratic Hamiltonian,
\[
\overline{H}^{(2)}=\int \frac 12\omega ^{\prime \prime }|\psi _x|^2dx+\int
\left[ \frac{\widetilde{\Psi }\widehat{k}\widetilde{\Psi }}{2\left( \rho
+1\right) }+(1-\rho )\frac{g\widetilde{\eta }^2}2\right] dx,
\]
where the first part relates to the dispersion of the wave packet and the
second one describes long linear gravity waves. In this case, the total
Hamiltonian is the sum of these two terms:
\[
\overline{H}_{\mathrm{LF}}=\overline{H}^{(2)}+\overline{H}^{(3)}.
\]
Note that $\overline{H}_{\mathrm{LF}}$ (subscript $LF$ means low frequency)
is only one part of the full Hamiltonian. It is used to find the coefficient
$\gamma $, which can be found independently.

The Hamiltonian equations of motion for $\overline{H}_{LF}$ are written
similarly to (\ref{canonical}) and (\ref{k-rep}):
\[
i\psi _t=\frac{\delta \overline{H}_{\mathrm{LF\;}}}{\delta \psi ^{*}}=-\frac
12\omega ^{\prime \prime }\psi _{xx}+\frac{Ak_0}{\left( \rho +1\right) }%
\widetilde{\Psi }_x\cdot \psi ,
\]
\[
\stackrel{\sim }{\eta }_t=\frac{\delta \overline{H}_{\mathrm{\;LF}}}{\delta
\widetilde{\Psi }}=\frac{\widehat{k}\widetilde{\Psi }}{\left( \rho +1\right)
}-\frac{Ak_0}{\left( \rho +1\right) }\frac{\partial |\psi |^2}{\partial x}%
,\quad \stackrel{\sim }{\Psi }_t=-\frac{\delta \overline{H}_{\mathrm{LF}}}{%
\delta \widetilde{\eta }}=-(1-\rho )g\widetilde{\eta }.
\]
Combining the last two equations gives
\[
-\frac 1{(1-\rho )g}\widetilde{\Psi }_{tt}=\frac{\widehat{k}\widetilde{\Psi }%
}{\left( \rho +1\right) }-\frac{Ak_0}{\left( \rho +1\right) }\frac{\partial
|\psi |^2}{\partial x}.
\]
The left-hand side of this equation can be neglected in comparison with the
right-hand side terms. Therefore
\[
\widetilde{\Psi }=Ak_0\widehat{k}^{-1}\frac{\partial |\psi |^2}{\partial x}.
\]
This leads to the following equation for $\psi $ :
\begin{equation}
i\psi _t+\frac 12\omega ^{\prime \prime }\psi _{xx}+\frac{A^2k_0^2}{\left(
\rho +1\right) }\widehat{k}|\psi |^2\cdot \psi =0.  \label{dysthe}
\end{equation}
In the context of gravity waves the nonlinear term in this equation was
first found by Dysthe ($A=1$) \cite{Dysthe}. It is of focusing type. This
corresponds to the second nonlocal contribution to the Hamiltonian:

\[
\overline{H}_{\mathrm{nonlocal}}^{(3)}=-\int \frac{k_0^2}{2\left( \rho
+1\right) }A^2|\psi |^2\widehat{k}|\psi |^2dx.
\]
Combined with (\ref{NL-4}) it gives the final form for the nonlocal
interaction Hamiltonian:
\begin{equation}
\overline{H}_{\mathrm{nonlocal}}=\ -\int \frac{k_0^2}{2\left( \rho +1\right)
}|\psi |^2\widehat{k}|\psi |^2dx<0.  \label{nonlocal}
\end{equation}
Hence the coefficient $\gamma $ is given by
\[
\gamma =\frac{k_0^2}{(1+\rho )}.
\]
This coefficient being multiplied by $-1$, it corresponds to an attraction
between waves (focusing nonlinearity).

Let us now consider the coefficient $\beta $, which is responsible for the
local four-wave interactions (\ref{wave4}). It is possible to develop the
same scheme as the one we used in calculating the coefficient $\gamma $.
Another way is to calculate the first derivative of the matrix element $%
\widetilde T_{12|34}$ with respect to its arguments at the wavenumber $%
k_i=k_0$ (of course, excluding the nonlocal interaction which has been
defined already). We skip all these calculations and present directly the
final answer:
\[
\beta =\frac{3k_0^2}{16(1+\rho )},
\]
so that
\[
\gamma =\frac{16}{3}\beta .
\]
The six-wave coupling coefficient $C$ is more complicated to compute. The
procedure requires applying the diagram technique (which, in fact, is a
usual perturbation theory based on the use of multi-scale expansion
methods). The corresponding calculations are presented in Appendix A. They
give
\[
C=\frac{M k_0^6}{3\omega _0},
\]
where $M=\frac{289(21+8\sqrt{5})}{16384}\approx 0.685961$. Thus, the
six-wave interaction term also leads to a focusing nonlinearity.

\section{Solitary wave solutions}

\setcounter{equation}{0}

In the previous section, we computed the coefficients that are needed to
analyze solitary wave solutions and their bifurcations near the critical
density ratio. As pointed out in Section II, solitary wave solutions
represent stationary points of the Hamiltonian for fixed momentum $P$. When
the critical velocity is approached, the solitons are transformed into
envelope solitons. The envelope solitons, like the original ones, also
represent stationary points but of the mean Hamiltonian, averaged with
respect to fast oscillations,
\begin{equation}
\overline{H}=\int \left[ \frac{\omega _0}{4k_0^2}|\psi _x|^2+\frac{\mu}{2}
|\psi |^4+i\beta (\psi _x^{*}\psi -\psi _x\psi ^{*})|\psi |^2-\frac{\gamma}{2%
} |\psi |^2\widehat k|\psi |^2-C|\psi |^6\right] dx,  \label{averH}
\end{equation}
for fixed momentum. After averaging the momentum becomes the number of waves
$N$ $=\int |\psi |^2dx$ multiplied by $k_0$. The variational problem (\ref
{var2}) for the envelope solitons is then written as follows:
\begin{equation}
\delta (\overline{H}+\lambda ^2\omega _0N)=0.  \label{var3}
\end{equation}
The soliton shape in this case is governed by the corresponding
Lagrange-Euler equation:
\begin{equation}
-\lambda ^2\omega _0\psi +\frac{\omega _0}{4k_0^2}\psi _{xx}-\mu |\psi
|^2\psi +4i\beta |\psi |^2\psi _x+\gamma \psi \widehat k|\psi |^2+3C|\psi
|^4\psi =0.  \label{soliton1}
\end{equation}
It is important to note that all terms in (\ref{averH}) are small compared
with $\omega _0N$. However, in the soliton solution both dispersion and
nonlinearity occur at a level comparable with $\lambda ^2\omega _0N$, where $%
\lambda ^2=(V_{cr}-V)/V_{cr}\ll 1$ . It follows, for instance, from the
variational problem (\ref{var3}) or from the equation (\ref{soliton1}) for
the soliton shape.

Equation (\ref{soliton1}) is a pseudo-differential equation. Besides
differentiation it contains the integral operator $\widehat k$ that
introduces some complexity in its analysis. The equation (\ref{soliton1})
can be simplified by introducing the amplitude $r$ and the phase $\varphi$
of $\psi$. Substituting $\psi =re^{i\varphi }$ into Eq. (\ref{soliton1}) and
separating real and imaginary parts give the following equation for the
phase:
\begin{equation}  \label{phase-eq}
\varphi _x=-\beta \frac{4k_0^2}{\omega _0}r^2.
\end{equation}
Incidentally, the phase equation for the classical NLSE (\ref{DNLS}) is
\[
\varphi_x r^2 = \mbox{constant}.
\]
Since $r \to 0$ as $|x| \to \infty$, the constant is equal to zero and there
is no dependence on $x$ of the phase as opposed to the present case. In
nonlinear optics, the dependence on $x$ of the phase $\varphi$ is called
pulse chirp. By means of the relation (\ref{phase-eq}), the phase is
excluded from the equation for the amplitude $r$:
\begin{equation}
-\lambda ^2\omega _0r+\frac{\omega _0}{4k_0^2}r_{xx}-\mu r^3+\gamma
r\widehat k(r^2)+3C_1r^5=0,  \label{eq=r}
\end{equation}
where the six-wave coupling coefficient $C$ is renormalized as
\[
C_1=C+\frac{4k_0^2}{\omega _0}\beta ^2.
\]
Substituting the rescaling
\begin{equation}
x=x^{\prime }\frac{\sqrt{3\omega _0C_1}}{2k_0|\mu |},\quad r=r^{\prime }%
\sqrt{\frac{|\mu |}{3C_1}}, \quad \lambda ^{^{\prime }}=\frac \lambda {|\mu
|}\sqrt{3C_1\omega _0},\quad \gamma ^{^{\prime }}=\gamma \frac{2k_0}{\sqrt{%
3\omega _0C_1}}=\frac{32}{\sqrt{397}}\approx 1.60603  \label{scaling}
\end{equation}
in Eq. (\ref{eq=r}) yields
\begin{equation}
-\lambda ^2r+r_{xx}-\mu r^3+r^5+\gamma r\widehat k(r^2)=0,  \label{norm-r}
\end{equation}
where all primes have been omitted. The value of $\mu$ is $\mu =-1$ for $%
\rho <\rho_{cr}$ and $\mu =1$ for $\rho >\rho_{cr}$.

From the scaling (\ref{scaling}) it is seen that the new $\lambda ^2$ takes
small values when the old $\lambda $ goes to $0$, and, respectively, large
values when the old $|\mu |$ approaches $0$ . Below the critical density
ratio ($\mu =-1$) all nonlinear terms are focusing and, thus, in this case
soliton solutions exist in the whole range of (new) $\lambda ^2$. For small $%
\lambda ^2$, the last two terms in Eq. (\ref{norm-r}) can be neglected and
solitons are those of the classical NLSE:
\begin{equation}
r=\frac{\sqrt{2}\lambda }{\cosh (\lambda x)}.  \label{NLS-soliton}
\end{equation}
Thus, at the critical velocity $V_{cr}$, solitons with $\mu <0$ undergo
supercritical bifurcation: their amplitude vanishes like $\left(
1-V/V_{cr}\right) ^{1/2}$. When $\lambda $ increases, the soliton width
decreases while the soliton amplitude grows. This behavior is confirmed with
the numerical computation of solutions to this equation (see below). This
property can also be observed with $\gamma =0$ (i.e. in the absence of
nonlocal nonlinearity). Then this equation admits the following analytical
solution expressed in terms of elementary functions \cite{DI96},\cite{K99}:
\begin{eqnarray}
r^2 &=&\frac{4\lambda ^2}{\sqrt{1+16\lambda ^2/3}\,\cdot \cosh (2\lambda
x)-\mu },  \label{r-soliton} \\
\varphi  &=&-\frac{\beta ^2}{\sqrt{C_1}}\tan ^{-1}\left[ \frac{\sqrt{%
1+16\lambda ^2/3}\cdot e^{2\lambda x}\,-\mu }{4\lambda /\sqrt{3}}\right] .
\end{eqnarray}
This solution is valid for both signs of $\mu $. It is seen that this
solution approaches the NLS soliton (\ref{NLS-soliton}) when $\lambda $ goes
to zero, if $\mu =-1$ (focusing nonlinearity). For $\mu =1$ the solution (%
\ref{r-soliton}) at $\lambda =0$ transforms into a soliton with algebraic
decay at infinity:
\begin{equation}
r^2=\frac 2{x^2+4/3}.  \label{power}
\end{equation}
Such solitons have been studied in \cite{I97}.

If $\gamma \neq 0$, the soliton solution decays exponentially as $%
|x|\rightarrow \infty $ for both signs of $\mu$ ($\psi \propto e^{-\lambda
|x|}$), except for $\lambda =0$. In this case, with $\mu =1$ (defocusing
nonlinearity), an asymptotic analysis of Eq. (\ref{norm-r}) shows that the
soliton solution will decay algebraically, similarly to (\ref{power}):
\begin{equation}
r \propto \left(2 - \gamma\frac{N_{0}}{\pi }\right) ^{1/2}\frac{1}{|x|},
\label{asymp}
\end{equation}
where $N_{0}$ is the number of waves on this soliton solution. This
asymptotic behavior suggests to seek for solutions in a form similar to (\ref
{power}):
\begin{equation}
r^2=\frac{A^2}{x^2+a^2},  \label{r-power}
\end{equation}
where $a$ and $A$ are unknown constants. Substituting this ansatz into Eq. (%
\ref{norm-r}) and assuming $\lambda =0$ yields
\begin{eqnarray*}
\frac{r_{xx}}r & = & \frac 2{x^2+a^2}-\frac{3a^2}{\left( x^2+a^2\right) ^2},
\\
-\gamma \frac \partial {\partial x}\widehat{H}(r^2) & = & -\frac{\gamma A^2}%
a\left( \frac 1{x^2+a^2}-\frac{2a^2}{\left( x^2+a^2\right) ^2}\right), \quad %
\mbox{since} \;\; \widehat{H}\left( \frac 1{x^2+a^2}\right)=-\frac x{a\left(
x^2+a^2\right)}, \\
-r^2+r^4 &= & -\frac{A^2}{x^2+a^2}+\frac{A^4}{\left( x^2+a^2\right) ^2}.
\end{eqnarray*}

Setting equal terms that are proportional to $\left( x^2+a^2\right) ^{-1}$
and $\left( x^2+a^2\right) ^{-2}$, one obtains two equations for determining
$a$ and $A$:
\[
2-A^2-\frac{\gamma A^2}a=0,\quad -3a^2+A^4+2a\gamma A^2=0.
\]
The solution of this system with $a>0$ is
\begin{equation}
a=\frac{1}{3}\left(-\gamma+2\sqrt{\gamma^{2}+3}\right),\quad
A^2=\gamma^{2}+2-\gamma\sqrt{\gamma^{2}+3}.  \label{power-par}
\end{equation}
For this solution the number of waves is given by the expression
\begin{equation}
N_{0}=\pi \left(\sqrt{\gamma^2+3}-\gamma\right)\approx 2.37514.
\label{crit-N}
\end{equation}
This is the critical soliton solution with finite amplitude at $\lambda =0$.
In other words, solitons above the critical density ratio ($\mu >0$) undergo
a subcritical bifurcation. The numerical value of $N_{0}$ is $2.38995$,
which is good agreement with (\ref{crit-N}).

When the density ratio is equal to the critical one ($\mu =0$), equation (%
\ref{eq=r}) admits another rescaling, different from (\ref{scaling}):
\[
\widetilde{x}=2\lambda k_0 x, \quad \widetilde{r}=r\sqrt{\frac{\lambda
^2\omega _0}{3C_1}}.
\]
Under this new scaling equation (\ref{eq=r}) becomes
\begin{equation}
-\lambda ^2r+r_{xx}+\gamma r\widehat kr^2+r^5=0,  \label{critNLS}
\end{equation}
where tildes have been omitted and $\gamma$ is given by the last relation in
(\ref{scaling}). Eq. (\ref{critNLS}) can be related to the critical
stationary NLS equation. When $\gamma =0$ this equation is nothing else than
the quintic NLS equation which, as is well-known (see, for instance, \cite
{kuz-tur} or the review \cite{KRZ}), belongs to the class of critical NLS
equations. However, this property persists in the presence of nonlocal
nonlinearity. It follows, for instance, if one considers the behavior of the
number of waves $N$ on the parameter $\lambda ^2$ for the soliton solution.
Indeed, the change
\begin{equation}
r=\sqrt{\lambda }g(\lambda x), \quad \xi=\lambda x,  \label{scaling-crit}
\end{equation}
reduces Eq. (\ref{critNLS}) to a form not containing $\lambda$:
\begin{equation}
-g+g_{\xi \xi }+\gamma g\widehat k_\xi (g^2)+g^5=0.  \label{critNLS1}
\end{equation}
Consequently, the number of waves for soliton solutions is independent on $%
\lambda $:
\[
N=N_{cr}=\int g^2(\xi )d\xi .
\]

As $\lambda \rightarrow \infty $, we also come up asymptotically with the
same equation (\ref{critNLS}). In this limit, and independently on $\mu ,$
the main balance of the term $-\lambda ^2r$ in equation (\ref{norm-r}) comes
from the last two last terms in its left-hand side. The latter means that we
have convergence of two soliton branches corresponding to different signs of
$\mu $ as $\lambda \rightarrow \infty $ .

It is easy to show also that the Hamiltonian evaluated on the soliton
solution of (\ref{critNLS}) is equal to zero. However, for solitons with $%
\mu >0$, the Hamiltonian takes positive values, while for $\mu <0$ (focusing
case) the Hamiltonian of solitons happens to be negative. This is a simple
consequence of the variational problem (\ref{var3}).

Indeed, consider the scaling transformation that keeps the number of waves $N
$ unchanged:
\begin{equation}
r\rightarrow \frac 1{a^{1/2}}r\left( \frac xa\right) ,  \label{scaling-r}
\end{equation}
where $r$ is a soliton solution of equation (\ref{norm-r}). Under this
transform the Hamiltonian $\overline{H}$ (\ref{averH}) becomes a function of
the scaling parameter $a$:

\begin{equation}
\overline{H}=\frac 1{a^2}\int \left( r_x^2-\frac{\gamma}{2}r^2\widehat
kr^2-\frac 13r^6\right) dx+\frac 1a\int \frac{\mu}{2} r^4dx.
\label{scalingH}
\end{equation}
Because the soliton solution is a stationary point of $\overline{H}$ (\ref
{var3}),
\[
\left. \frac{\partial \overline{H}}{\partial a}\right| _{a=1}=0,
\]
and therefore
\[
\int \left( r_x^2-\frac{\gamma}{2}r^2\widehat kr^2-\frac 13r^6\right)
dx=-\frac 14\int \mu r^4dx.
\]
Hence we have
\[
\overline{H}_s=\frac 14\int \mu r^4dx.
\]
Thus, at $\mu =0$, $\overline{H}_s=0$. If $\mu <0$, $\overline{H}_s<0$ and $%
\overline{H}_s>0$ if $\mu >0$.

Solitons of both equations (\ref{norm-r}) and (\ref{critNLS}) with different
values of $\lambda $ were found numerically for both signs of $\mu $ as well
as for $\mu =0$. In order to find them, we used the method suggested by V.I.
Petviashvili \cite{petviashvili}. The idea of the Petviashvili method is
based on the solution of diffusive type equations with the introduction of a
new time $\tau $. Stationary solutions of this extended equation coincide
with the soliton solution of the original equation that we are looking for.
In the present case we developed the following algorithm.

The iteration scheme to find solutions connects values $r^{(n)}(x)$ at the $n
$th time step with $r^{(n+1)}$ by the relation:
\begin{equation}
r^{(n+1)}=\left\{ r+\triangle \tau \cdot \left[ -\lambda ^2r+r_{xx}-\mu
r^3+r^5+\gamma r\widehat k(r^2)\right] \right\} ^{(n)}M\left\{
r^{(n)}\right\} .  \label{iteration}
\end{equation}
Here $\triangle \tau $ is the time step and $M\left\{ r^{(n)}\right\} $ is
the functional of $r^{(n)}$ of the form
\[
M\left\{ r\right\} =\left[ \frac{||-\lambda ^2r+r_{xx}-\mu r^3+\gamma
r\widehat k(r^2)||_{L_2}}{||r^5||_{L_2}}\right] ^{1/8},
\]
where $||f||_{L_2}$ is the $L_2$-norm. It is seen from this equation that
its stationary solution satisfies the SNLS equation (\ref{norm-r}). The
presence of the norm $||r^5||_{L_2}$ in the denominator of $M\left\{
r\right\} $ provides attraction to the nontrivial solution of (\ref{norm-r}%
), with $r\neq 0.$ The efficiency of this scheme has been demonstrated and
it provides very fast convergence. The iteration scheme (\ref{iteration})
works very well for $\mu >0$ and $\mu =0$. For $\mu <0,$ however, it was
convenient to use in (\ref{iteration}) the factor $M\left\{ r\right\} $ in
the form:
\[
M\left\{ r\right\} =\left[ \frac{||-\lambda ^2r+r_{xx}+\gamma r\widehat
k(r^2)||_{L_2}}{||-\mu r^3+r^5||_{L_2}}\right] ^{1/4}.
\]
Figures 1 and 2 show typical soliton shapes for $\mu =\pm 1$ when $\lambda
\neq 0$. Numerically we checked that solitons decay at infinity like $%
e^{-\lambda |x|}$.
\begin{figure}[tbp]
\includegraphics[width=12cm,height=8cm]{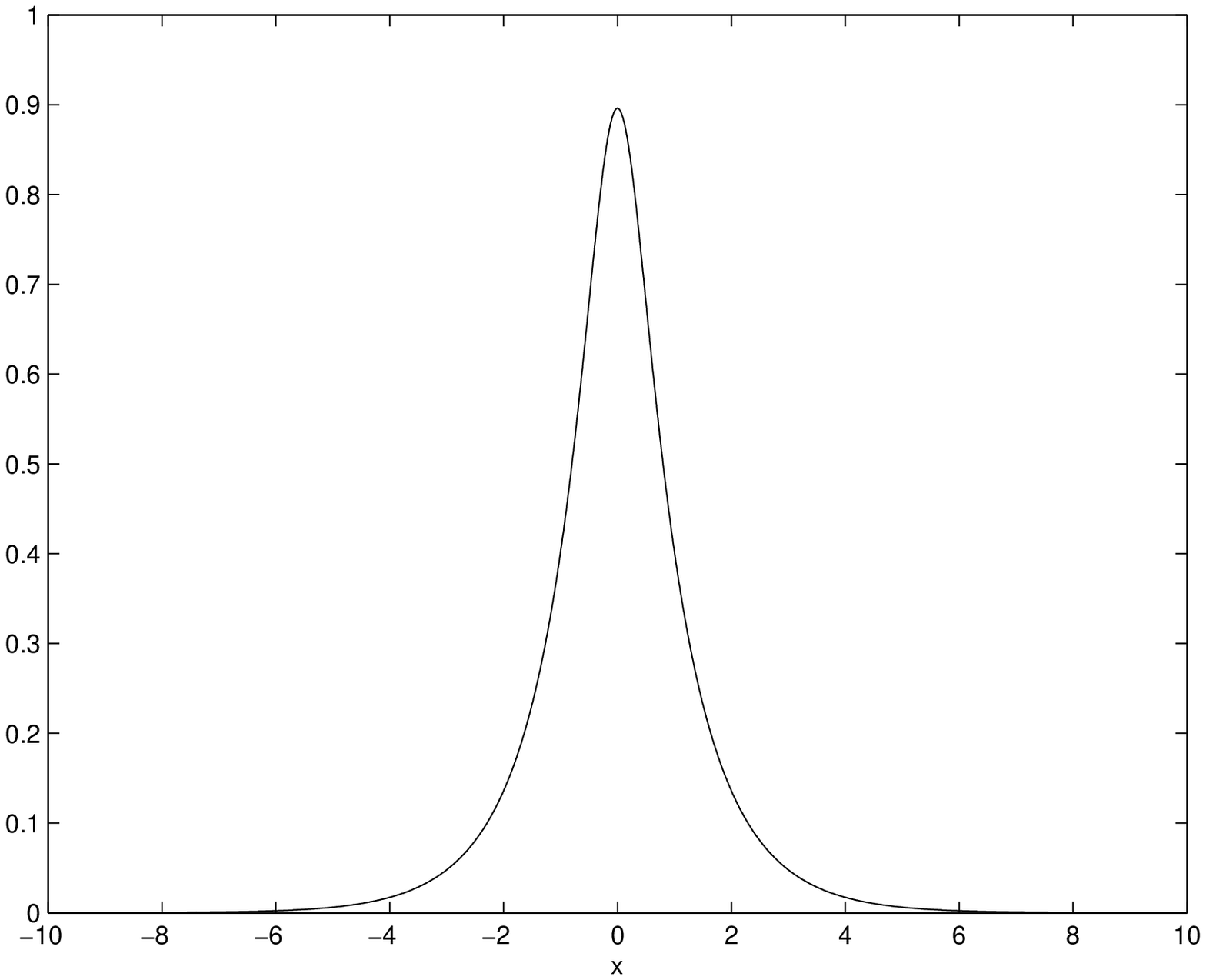}
\caption{Soliton shape $r(x)$ with parameters $\mu =-1$, $\lambda =1$.}
\end{figure}

\begin{figure}[tbp]
\includegraphics[width=12cm,height=8cm]{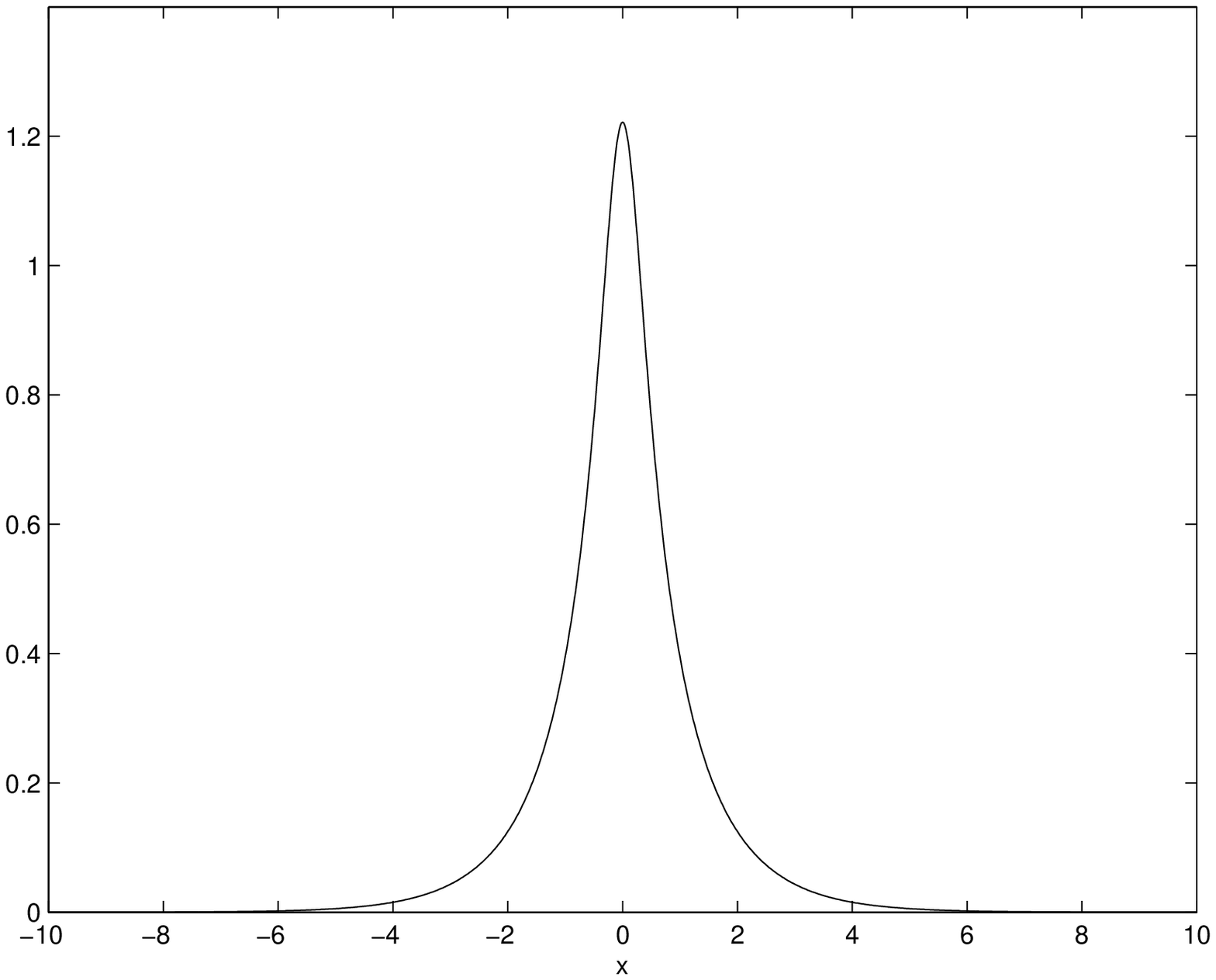}
\caption{Soliton shape $r(x)$ with parameters $\mu =1$, $\lambda =1$.}
\end{figure}

For $\lambda =0$ and $\mu =1$ the iteration scheme gave the soliton
dependence (Fig. 3), which is in agreement with the analytical result (\ref
{r-power},\ref{power-par}) at very high accuracy (for example, the factor $M$
at the stationary solution, presented in the paper, was equal to 1 within an
error less than $10^{-5}$). The time step $\Delta \tau $ was equal to $%
0.00018$. All derivatives in the equations, as well as the action of the
integral operator $\widehat{k}$, were calculated by means of the standard
FFT program. The initial conditions for the iteration procedure (\ref
{iteration}) were taken in the form of solitons (\ref{r-soliton}) with $%
\gamma =0$, which provide exponential decay while approaching the ends of
the numerical interval on $x$. We chose a symmetric numerical domain $[-a,a]$%
, with $a=10$ for all runs except the case $\lambda =0$ where we took $a=50.$
The soliton shape in this particular case is presented in Fig. 3.
\begin{figure}[tbp]
\includegraphics[width=12cm,height=8cm]{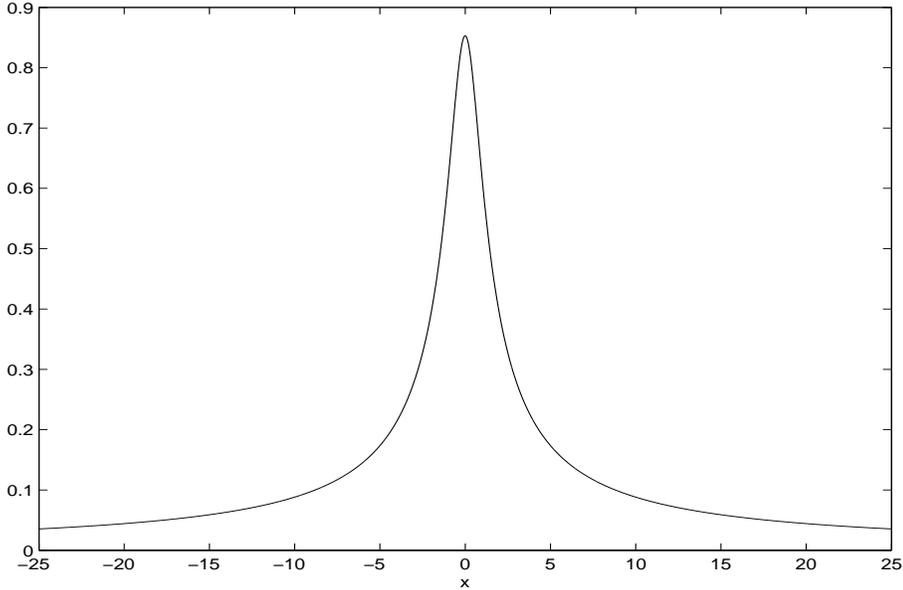}
\caption{Soliton shape $r(x)$ with parameters $\mu =1$, $\lambda =0$.}
\end{figure}
Fig. 4 shows the dependence of $N$ on the soliton parameter $\lambda $. It
is seen that the lower soliton branch and the upper soliton branch converge
at large $\lambda $. Between the upper and the lower curves we have a
straight line corresponding to the critical solitons with $N=N_{cr}$.
\begin{figure}[tbp]
\includegraphics[width=12cm,height=8cm]{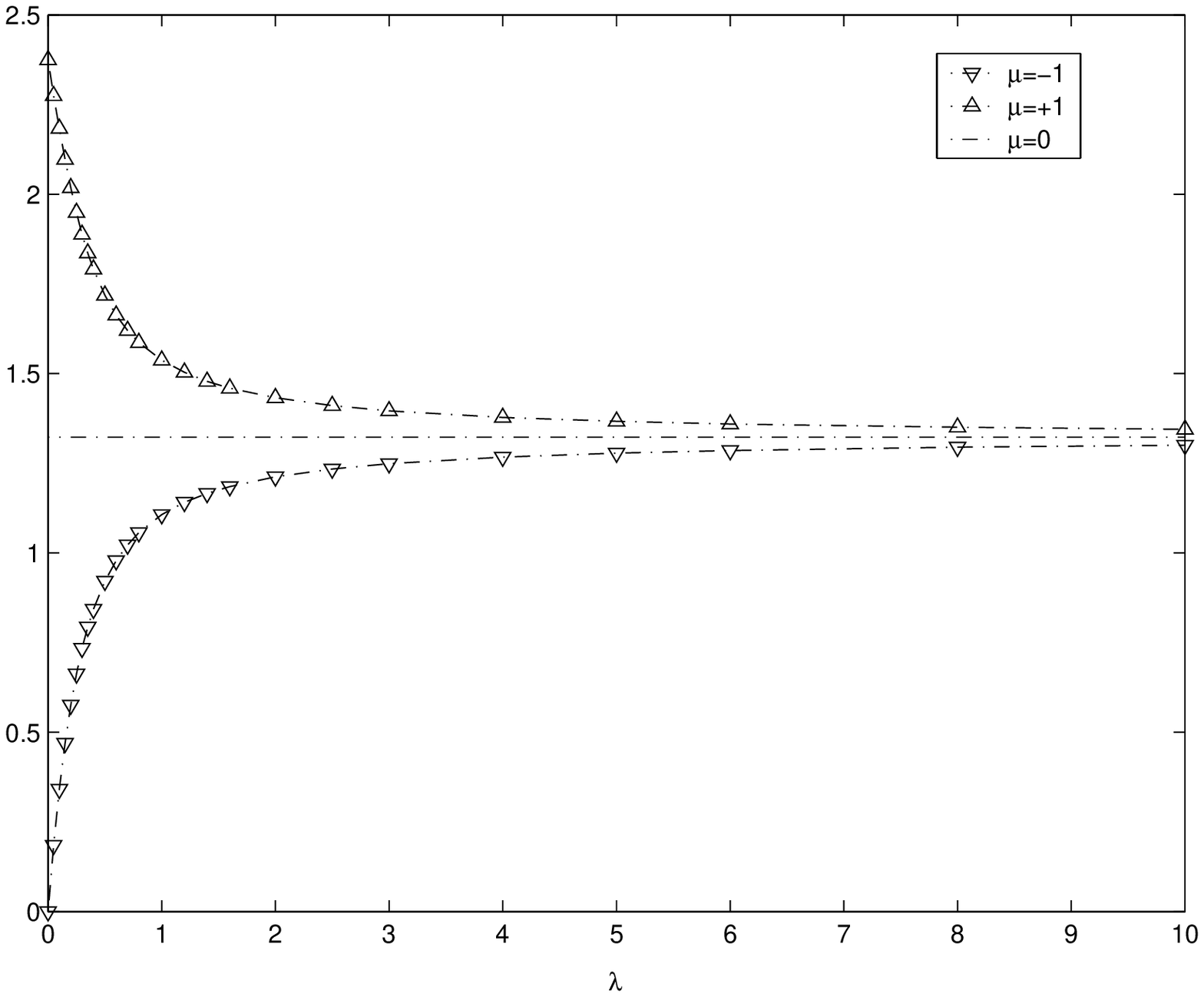}
\caption{Dependence of $N$ on $\lambda $ for $\mu =\pm 1,0$}
\end{figure}
The same separation takes place for soliton amplitudes (Fig. 5). At the
point $\lambda =0$ solitons undergo bifurcations.
\begin{figure}[tbp]
\includegraphics[width=12cm,height=8cm]{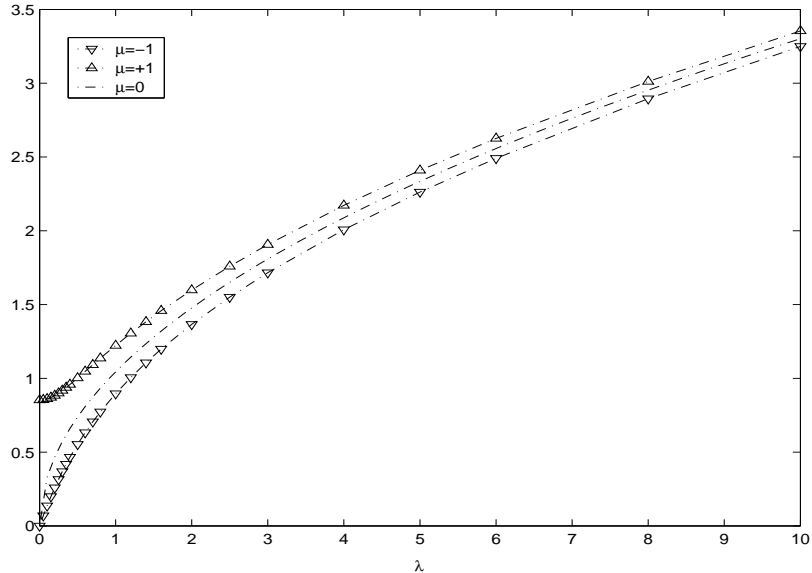}
\caption{Dependence of soliton amplitude on $\lambda $}
\end{figure}
\noindent
For negative $\mu $ we have a supercritical soliton bifurcation, while a
subcritical occurs when $\mu >0$, with the soliton amplitude jump given by (%
\ref{power-par}).

\section{Stability of solitons}

\setcounter{equation}{0} In this section we study the stability of the
solitons that we analyzed in the previous section. In order to do that, we
use the method based on Lyapunov's theorem. Applying this theorem to the
present Hamiltonian system means the following: a soliton considered as a
stationary point of the Hamiltonian $\overline{H}$ for a fixed number of
waves $N$ is stable if it realizes a minimum (or maximum) of the
Hamiltonian. If a stationary point represents a saddle point then the
corresponding soliton is expected to be unstable. The latter is only an
indication that solitons are unstable. For instance, the classical
counterexample is the Hamiltonian $H=p_1^2/2+q_1^2/2-p_2^2/2-q_2^2/2$ for a
system with two degrees of freedom where the saddle point $p=q=0$ is stable.

In fact this type of indication is already available if one looks at the
scaling (\ref{scaling-r},\ref{scalingH}) for which
\[
\overline{H}\left( a\right) =\left( \frac 1a-\frac 1{2a^2}\right) \frac{\mu}{%
2} \int r_s^4 \, dx,
\]
where $r_s$ is the solitary wave solution. Hence, it is seen that for
negative $\mu $ the function $\overline{H}\left( a\right) $ is bounded from
below and its minimum, $\overline{H}_s<0,$ corresponds to the soliton. On
the contrary, for $\mu >0$, this function has a maximum equal to $\overline{H%
}_s>0$, which is unbounded from below as $a\to 0$. Moreover, in the latter
case, it is possible to see that the stationary point corresponding to the
soliton solution is a saddle point. It follows immediately if, in addition
to the scaling transformation, one considers the gauge transformation $\psi
_s\rightarrow \psi _se^{i\chi }$ under which
\[
\overline{H}\to \overline{H}_s+\int r^2\left( \chi _x\right) ^2dx.
\]
For $\mu <0$, however, this transformation shows that the soliton solution
remains the minimum point. We reach the same conclusion if we apply the
Vakhitov-Kolokolov criterion \cite{VK} to the soliton solutions. The
criterion states that, if
\[
\frac{\partial N_s}{\partial \lambda ^2}<0,
\]
then solitons are unstable and they are stable in the opposite case. This
criterion has a simple physical interpretation. The quantity $-\lambda ^2$
is related to the energy of the soliton as a bound state. If by adding one
particle (i.e. increasing $N$) this level shifts towards the continuous
spectrum, then obviously such bound state will be unstable. As we saw in the
previous section the derivative $\partial N_s/\partial \lambda ^2$ is
negative for $\mu >0$ and becomes positive when $\mu <0$ . We would like to
emphasize that the Vakhitov-Kolokolov criterion was derived for the
classical NLS equation and, strictly speaking, cannot be applied to our
system. Thus, we have again indications that support stability of the lower
soliton branch and, respectively, instability for the upper branch. Note
that this stability indication is consistent with that for the cubic NLS
solitons (\ref{NLS-soliton}) which, as is well-known, are stable (see, e.g.
\cite{KRZ}).

Now we show that the lower soliton branch ($\mu<0$) is stable: solitons from
this branch indeed realize a minimum of the Hamiltonian $\overline{H}$ for a
fixed number of waves $N$. The dimensionless Hamiltonian $\overline{H}$
written in terms of amplitude and phase reads
\begin{equation}
\overline{H}=\int \left[ r_x{}^2+\frac{\mu}{2} r^4 - \frac{\gamma}{2}
r^2\widehat kr^2-\frac 13r^6+r^2\left( \varphi _x+\beta r^2\right) ^2\right]
dx.  \label{Ham1}
\end{equation}
The last term here is positive definite and vanishes exactly on the
stationary solitons. Thus, we now should prove that the minimum of $%
\overline{H}$ is reached on the soliton obtained as solution of Eq. (\ref
{norm-r}).

Consider the integral
\[
I=\int \left[ \frac{\gamma}{2}r^2\widehat kr^2+\frac 13r^6\right] dx
\]
and get its estimate through the integrals $\int r_x{}^2dx$ and $N$. As was
shown in \cite{K99}, we have the following estimate in the absence of
nonlocal terms:
\[
\int r^6dx\leq \left( \frac N{N_1}\right) ^2\int r_x{}^2 \, dx,
\]
where $N_1=\int \psi _0{}^2dx$ and $\psi _0$ is the soliton solution for the
quintic (critical) NLS equation,
\[
-\psi _0+\psi _{0xx}+3\psi _0^5=0.
\]
This solution can be easily found and gives $N_1=\pi /2$.

For the integral $I_k=\int r^2\widehat kr^2dx$, we can write the following
set of inequalities:
\begin{eqnarray*}
\int r^2\widehat kr^2dx &\leq &\max_x\left( r^2\right) \int r\widehat
kr\,dx\leq \int_{-\infty }^{x_{\max }}\left( r^2\right) _x\,dx\left( \int
r^2dx\right) ^{1/2}\left( \int r\widehat k^2r\,dx\right) ^{1/2} \\
&\leq &C_2N\int r_x{}^2\,dx.
\end{eqnarray*}
This inequality can be made sharper. To do that, consider the functional
\[
F\left\{ r\right\} =\frac{\int r^2\widehat kr^2dx}{\int r^2dx\cdot \int
r_x{}^2dx}.
\]
Its minimum value will give the best constant $C_2$. To find it one needs to
determine a minimizer among all stationary points of the functional $%
F\left\{ r\right\} $ . The stationary points of $F\left\{ r\right\} $ are
defined from the solutions of the equation
\[
2r_0\widehat kr_0^2-r_0+r_{0xx}=0.
\]
The minimizer for $F\left\{ r\right\} $ is given by the ground soliton
solution of this equation. It is a symmetric function without nodes. Hence
the best constant
\[
C_{2,\mathrm{best}}=\frac 1{N_2}, \quad N_{2}=\int r_0^2dx\approx 1.39035.
\]
Thus,
\begin{equation}
I\leq \left( \frac{\gamma}{2}\frac N{N_2}+\frac{N^2}{3N_1^2}\right) \int
r_x{}^2dx.  \label{ineq}
\end{equation}
This inequality allows to find the criterion for the Hamiltonian to be
bounded from below. Substituting (\ref{ineq}) into (\ref{Ham1}) for $\mu <0$
yields
\begin{equation}
\overline{H}\geq \left[ 1-\left( \frac{\gamma}{2}\frac N{N_2}+\frac{N^2}{%
3N_1^2}\right) \right] \int r_x{}^2dx-\frac{1}{2}\int r^4\,dx.  \label{ineq1}
\end{equation}
Hence it is seen that the Hamiltonian is bounded from below if
\[
\frac{\gamma}{2}\frac N{N_2}+\frac{N^2}{3N_1^2}\leq 1,
\]
or
\begin{equation}
N\leq N_3=\sqrt{\frac{9}{16}\gamma^2\frac{N_{1}^{4}}{N_2^2}+3N_1^2}-\frac{3}{%
4}\gamma \frac{N_1^2}{N_2}.  \label{ineq2}
\end{equation}
The final step in proving the Hamiltonian boundedness is based on the
estimate for the integral $\int r^4dx$. According to \cite{ZK98}
\[
\int r^4dx\leq \frac 1{\sqrt{3}}N^{3/2}\left( \int r_x{}^2dx\right) ^{1/2}.
\]
Substitution of this inequality into the estimate (\ref{ineq1}) results in
the desired bound for $\overline{H}$:
\[
\overline{H}\geq -\frac{N^3}{4\sqrt{3}}\left[ 1-\left( \frac{\gamma}{2}\frac
N{N_2}+\frac{N^2}{3N_1^2}\right) \right] .
\]
It turns out that the numerical value of $N_3=1.3224$ is almost the same as
the critical number $N_{cr}=1.3225$, defined by solitons with $\mu =0$.
Thus, solitons from the lower branch which satisfy the criterion (\ref{ineq2}%
) are stable not only with respect to small perturbations but also against
finite ones. Concerning the solitons from the upper branch, they are all
unstable with respect to finite disturbances.

In summary, we would like to emphasize once more  the difference between $%
N_3$ and $N_{cr}$ although the derivative $\partial N_s/\partial \lambda ^2$
is positive for the whole lower soliton branch and according to the
Vakhitov-Kolokolov criterion this branch is expected to be stable. Up to now
it is an open question but we think that it is so and $N_3$ and $N_{cr}$
should coincide.

\section{Concluding remarks}

Thus, we have analyzed the behavior of solitons near the critical density
ratio $\rho =\rho _{cr}$. Above $\rho _{cr}$ solitons undergo a subcritical
bifurcation with the amplitude jump proportional to $\sqrt{\rho -\rho _{cr}}$%
. Therefore our theory based on the Hamiltonian expansion works when this
jump is small, i.e. in a small vicinity of $\rho =\rho _{cr}$. In order to
describe solitons a generalized NLS equation is derived based on the
Hamiltonian average over fast oscillations with the carrying frequency
corresponding to the critical soliton velocity $V=V_{cr}.$ The generalized
NLS equation contains three kinds of nonlinear terms. The first one is
nothing else than the nonlinear frequency shift taking into account six-wave
nonlinear interactions. This is the local term. Another nonlinearity is
responsible for steepening the envelope solitons (this is the so-called
Lifshitz term \cite{LL}). And finally we have found the nonlocal
contribution familiar to that calculated by Dysthe for gravity waves. This
Dysthe term is focusing as well as the six-wave nonlinear interaction term.
Within the generalized NLS equation we analyzed the stability of solitons.
In particular, we have found a region of wave intensity, $N\leq N_3$, where
solitons are stable. Their stability is based on the boundedness of the
Hamiltonian from below. For solitons above $\rho _{cr}$ we have shown their
possible instability, at least, instability with respect to finite
perturbations. An interesting problem is the nonlinear stage of this
instability, whether it can lead to the collapse of solitons. The latter
question is important not only from the point of view of interfacial waves
but also because there exists some similarity between the generalized NLS
equation derived in this paper and those describing the behavior of short
optical pulses in fibers from the femtosecond range of pulse durations.

\section*{Acknowledgements}

The authors thank A.I. Dyachenko for valuable discussions concerning the
numerical aspects of this paper. Two authors (DA and EK) thank the Centre de
Math\'ematiques et de Leurs Applications (CMLA) of Ecole Normale
Sup\'erieure of Cachan, where this paper was initiated. This paper was
performed in the framework of the NATO Linkage Grant EST.CLG.978941. The
work of DA and EK was also supported by the Russian Foundation of Basic
Research and by the Program of Russian Academy of Sciences ``Mathematical
methods in nonlinear dynamics''.


\appendix

\section{Six-wave coupling coefficient}

\setcounter{equation}{0}

As pointed out in Section III, the calculation of the nonlinear coupling
coefficients $\beta $, $\gamma $ and of the six-wave coefficient $C$ can be
performed independently for each coefficient. Therefore for the coefficient $%
C$ this problem reduces to the calculation of the nonlinear frequency shift
for the main harmonic with $k=k_0$ and $\omega _0\equiv \omega (k_0)$ when
the first contribution to the nonlinear frequency shift proportional to $%
|\psi |^2$ is equal to zero. Second, in such a case, it is enough to
consider its limit of a monochromatic wave, instead of a quasi-monochromatic
wave. This means that one needs to develop the perturbation theory assuming
that
\begin{equation}
a_k=\sqrt{2\pi }\sum_nC_n(t)e^{-in\omega _0t}\delta (k-nk_0),  \label{app-1}
\end{equation}
where the leading order corresponds to the main harmonic $n=1$ and
amplitudes of all other (combined) harmonics are supposed to be small in
comparison with the first harmonic amplitude.

After this remark we introduce the small parameter $\epsilon $ so that
\begin{equation}
C_1=\epsilon \psi (T),  \label{app-2}
\end{equation}
where $T=\epsilon ^4t$ is a slow time and $\psi (T)$ is the envelope of the
main harmonic. Hence it is easy to see that all needed combined harmonics $%
C_n$ become functions of the slow time $T$ and take in their expansion
powers of the parameter $\epsilon $:
\begin{equation}  \label{app-3}
C_0=C_0^{*}=\epsilon ^2\psi _0+\epsilon ^4\psi _{01}, \quad C_{\pm
2}=\epsilon ^2\psi _{\pm 2}+\epsilon ^4\psi _{\pm 21}, \quad C_{-1}=\epsilon
^3\psi _{-1},\quad C_{\pm 3}=\epsilon ^3\psi _{\pm 3}.
\end{equation}
Note that in this case nonlinear interactions with the zeroth harmonic do
not give any contribution because the corresponding matrix elements vanish
by the same law as for three-wave interaction.

The equations of motion for amplitudes $C_n$ follow from the general
equation (\ref{k-rep}) for normal amplitude $a_k$:
\begin{equation}
\epsilon ^4\frac{\partial C_n}{\partial T}+i\left[ \omega (nk_0)-n\omega
_0\right] C_n=-i\frac{\partial \overline{H}_{\mathrm{int}}}{\partial C_n^{*}}%
.  \label{app-4}
\end{equation}
Here $\overline{H}_{\mathrm{int}}$ is mean value of the interaction
Hamiltonian after substitution expressions (\ref{app-1}) into (\ref{H-int})
and average. $\overline{H}_{\mathrm{int}}$ contains 19 matrix coefficients
needed to determine the constant $C$ :
\begin{eqnarray*}
&&\ U_{11-2},\smallskip\ U_{12-3},\smallskip\ V_{1|-12},\smallskip\
V_{1|-23},\smallskip\ V_{-1|1-2},\smallskip\ V_{2|11},\smallskip\ V_{3|12}, %
\smallskip\ V_{-2|1-3},\smallskip\ T_{11|11},\smallskip\ T_{11|20}, \\
& & T_{12|12},\smallskip\ T_{1-2|1-2},\smallskip\ T_{1|-111}^{(1,3)}, %
\smallskip\ T_{1|2-21}^{(1,3)},\smallskip\ T_{3|111}^{(1,3)},\smallskip\
T_{-3111}^{(0,4)},\smallskip\ T_{1|-2111}^{(1,4)},\smallskip\
T_{12|111}^{(2,3)},\smallskip\ T_{111|111}^{(3,3)},
\end{eqnarray*}
where $\pm 1\equiv \pm k_0$, $\pm 2\equiv \pm 2k_0$, and so on. These 19
coefficients are calculated in accordance with the rules explained in the
third section.

As is seen from the equation of motion (\ref{app-4}) the time derivative of $%
C_n$ is small compared with the second term $i\left[ \omega (nk_0)-n\omega
_0\right] C_n$, except the main harmonic where this term vanishes. In its
turn, the equation for $C_1$ contains terms proportional to $\epsilon ^3$
and $\epsilon ^5$. The terms of the third order ( $\sim $ $\epsilon ^3$) are
nothing more than the relation (\ref{crit-ratio}) serving for definition of
the critical Atwood number:
\[
T_{11|11}+\frac{2V_{2|11}^2}{2\omega _1-\omega _2}-\frac{2U_{11-2}^2}{%
2\omega _1+\omega _2}=0.
\]
The equation of motion for $\psi $ appears in the fifth order of $\epsilon $%
:
\begin{eqnarray*}
i\frac{\partial \psi }{\partial T} &=&2U_{11-2}\psi ^{*}\psi
_{-21}^{*}+2U_{12-3}\psi _2^{*}\psi _{-3}^{*}+2V_{1|-12}\psi _{-1}\psi
_2+2V_{1|-23}\psi _{-2}\psi _3 \\
&&\ +2V_{-1|1-2}\psi _{-1}\psi _{-2}^{*}+2V_{2|11}\psi _{21}\psi ^{*}\
+2V_{3|12}\psi _3\psi _2^{*}+2V_{-2|1-3}\psi _{-2}\psi _{-3}^{*} \\
&&\ +4T_{11|20}^{(2,2)}\psi ^{*}\psi _2\psi _0+4T_{12|12}^{(2,2)}\psi
_2^{*}\psi _2\psi +4T_{1-2|1-2}^{(2,2)}\psi _{-2}^{*}\psi _{-2}\psi
+3T_{1|-111}^{(1,3)}\psi _{-1}\psi \psi \\
&&\ +6T_{1|2-21}^{(1,3)}\psi _2\psi _{-2}\psi +6T_{11-1|1}^{(3,1)}\psi \psi
^{*}\psi _{-1}^{*}+6T_{12-2|1}^{(3,1)}\psi \psi _2^{*}\psi
_{-2}^{*}+3T_{3|111}^{(3,1)}\psi _3\psi ^{*}\psi ^{*} \\
&&\ \ \ \ +12T_{-3111}^{(0,4)}\psi _{-3}^{*}\psi ^{*}\psi
^{*}+4T_{1|-2111}^{(1,4)}\psi _{-2}\psi \psi \psi +12T_{-2111|1}^{(4,1)}\psi
\psi ^{*}\psi ^{*}\psi _{-2}^{*}\ \  \\
&&\ +2T_{12|111}^{(2,3)}\psi _2^{*}\psi \psi \psi +6T_{111|21}^{(3,2)}\psi
_2\psi \psi ^{*}\psi ^{*}\ \ +3T_{111|111}^{(3,3)}|\psi |^4\psi
\end{eqnarray*}
where all matrix elements are defined without factors of $\left( \sqrt{2\pi }%
\right) ^{2-n-m}$ as they had according to the definition of Section III
(compare with (\ref{U}), (\ref{V}), \ref{T-4})).

Here the amplitudes $\psi _0,\psi _{01},\psi _{\pm 2},\psi _{\pm 21},\psi
_{-1},\psi _{\pm 3}\ $ are found with the help of equations (\ref{app-4}).
This is a pure algebraic procedure which we performed with the help of
computer. As the result of substitution of $\psi _0,\psi _{01},\psi _{\pm
2},\psi _{\pm 21},\psi _{-1},\psi _{\pm 3}$ into the above equation we
arrive at the equation:
\[
i\frac{\partial \psi }{\partial T}=3\widetilde{T}_{111|111}^{(3,3)}|\psi
|^4\psi,
\]
where the six-wave coupling coefficient $C$ is positive:
\[
C=-\widetilde{T}_{111|111}^{(3,3)}=\frac{Mk_0^6}{3\omega _0},\smallskip\ M=%
\frac{289(21+8\sqrt{5})}{16384}\approx 0.685961.
\]
The corresponding nonlinear term is focusing.

\end{document}